\documentclass[11pt,reqno]{amsart}

\usepackage{packages}
\usepackage{notations}

\newcommand{\errorgraph}{
\begin{tikzpicture}[/pgf/declare function={f(\x)=gauss(\mean,\var);},scale=1.4,y=5cm]
\pgfdeclarepatternformonly{north east lines wide}%
   {\pgfqpoint{-1pt}{-1pt}}%
   {\pgfqpoint{10pt}{10pt}}%
   {\pgfqpoint{9pt}{9pt}}%
   {
     \pgfsetlinewidth{0.4pt}
     \pgfpathmoveto{\pgfqpoint{0pt}{0pt}}
     \pgfpathlineto{\pgfqpoint{9.1pt}{9.1pt}}
     \pgfusepath{stroke}
    }

\begin{axis}[
    height=6cm, width=12cm,
    no markers, domain={\xl}:{\xu },samples=100,
    axis lines=middle, xmax=\xu,
    xlabel=$x$, ylabel=\empty,
    xtick = \empty,
    ytick=\empty,
    xticklabels={$\mu$},
    enlargelimits=false, clip=false, axis on top,
    extra x ticks={\mean-7*2.4,\mean-5*2.4,\mean-3*2.4,\mean-2.4,\mean+2.4,\mean+2.4*3,\mean+2.4*5,\mean+2.4*7},
    extra x tick labels={$x_0$, $x_1$, $x_2$,$x_3$,$x_4$,$x_5$,$x_6$,$x_7$},
]

\addplot [name path=A,thick] {f(\x)};
\draw[<-,semithick,color=black] (\mean-4.8*2.2, 0.021) -- (\mean-6*2.4,0.03) node[above] {$f(x)$};

\addplot[name path=D,
    color=gray,
    fill=gray!40,
    opacity=.5,
    integral=\mean:\mean+3*\var - 4.8
] {f(\x)};
\addplot[name path=E,
    color=gray,
    fill=gray!40,
    opacity=.5,
    integral= \mean:\mean-3*\var+4.8
] {f(\x)};

\addplot[name path=D,
    color=gray,
    fill=yellow!20,
    opacity=.4,
    integral=\mean:\mean+3*\var - 4.8
] {1.2*f(\x)};
\addplot[name path=E,
    color=gray,
    fill=yellow!20,
    opacity=.4,
    integral= \mean:\mean-3*\var+4.8
] {1.2*f(\x)};
\draw[<-,semithick,color=purple] (\mean-0.5, 1.04*0.055) -- (\mean-1.2,1.1*0.06) node[above] {$\tilde{f}(x)$};

\draw[<-,semithick,color=brown,    opacity=1] (\mean+2, 0.052) -- (\mean+4,0.07) node[above] {$\mathrm{NoE}$};

\draw[<-,semithick,color=blue] (\mean+2.4+1.7, 0.047) -- (\mean+2.8*3,0.06) node[above] {$\mathrm{DE}$};
\draw[<-,semithick,color=blue] (\mean+2.4*3+1.7, 0.033) -- (\mean+2.8*3,0.06) node[above] {};
\draw[<-,semithick,color=blue] (\mean+2.4*5-1, 0.016) -- (\mean+2.8*3,0.06) node[above] {};
\draw[<-,semithick,color=blue] (\mean+2.4*7+1.7, 0.005) -- (\mean+2.8*3,0.06) node[above] {};

\draw[<-,semithick,color=green!40!black] (\mean+3*\var-4.8+0.5, 0.001) -- (\mean+3*\var-4.8+8,0.01) node[above] {$\mathrm{TE}$};

\draw[<-,semithick,color=magenta] (\mean+2.4*4.8+0.8, 0.007) -- (\mean+2.4*4.8+8,0.04) node[above] {$\mathrm{ThE}$};

\addplot[dashed,thick]
  coordinates {({\mean+3*\var-4.8},{5.8*f(\mean,\var)}) ({\mean+3*\var-4.8},{-2.07*f(\mean,\var)})}
  node[mydarkblue,below=-2pt]{$x_u+\frac{\Delta}{2}$}; 

\addplot[dashed,thick]
  coordinates {({\mean-3*\var+4.8},{5.8*f(\mean,\var)}) ({\mean-3*\var+4.8},{-2.07*f(\mean,\var)})}
  node[mydarkblue,below=-2pt]{$x_\ell-\frac{\Delta}{2}$}; 

\addplot[name path=C,magenta,dashed,thick]
      coordinates {({\mean+2.4*5+1.5},{1.8*f(\mean,\var)}) ({\mean+2.4*5+1.5},{-1.07*f(\mean,\var)})}
  node[mydarkblue,below=-2pt]{$x_{th}$}; 
  
\addplot[name path=B,magenta,dashed,thick]
      coordinates {({\mean+1.9*5},{1.6*f(\mean,\var)}) ({\mean+1.9*5},{-0.6*f(\mean,\var)})}
  node[mydarkblue,below=-2pt]{};

\addplot[name path=BD, mark=none, black, opacity=0, samples=2] {0.0195};

    \path[name path=xaxis]
      (0,0) -- (\pgfkeysvalueof{/pgfplots/xmax},0); 
      
      \addplot[pattern={north east lines},pattern color=black] fill between[of=xaxis and BD, soft clip={domain=\mean+1.9*5:\mean+2.4*5+1.5}];

    \path[name path=xaxis]
      (\xl,0) -- (\pgfkeysvalueof{/pgfplots/xmax},0);
    \addplot[green!40!black] fill between[of=xaxis and A, soft clip={domain= {\mean+3*\var-4.8}: \mean+3*\var}];
    \addplot[green!40!black] fill between[of=xaxis and A, soft clip={domain= {\mean-3*\var}: \mean-3*\var+4.8}];

\end{axis}

\end{tikzpicture}
}

\date{\today}
\title{Quantum Monte Carlo Integration \\ for Simulation-Based Optimisation}
\makeatletter
\let\inserttitle\@title
\makeatother

\author{Jingjing Cui~$^{2}$,
Philippe J.S. de Brouwer~$^{3}$,
Steven Herbert$~^{1}$, 
Philip Intallura~$^{4}$,
Cahit Kargi~$^{1}$,
Georgios Korpas~$^{5,6,7}$,
Alexandre Krajenbrink~$^{2}$,
William Shoosmith~$^{4}$,
Ifan Williams~$^{1}$,
Ban Zheng~$^{8}$}

\address{~$^{1}$ Quantinuum, Terrington House, 13–15 Hills Road, Cambridge CB2 1NL, United Kingdom}

\address{~$^{2}$ Quantinuum, Partnership House, Carlisle Place, London SW1P 1BX, United Kingdom}

\address{~$^{3}$ HSBC Service Delivery Sp. z o.o., Kapelanka 42A, Krakow, Poland}

\address{~$^{4}$ Quantum Technologies Group, Emerging Technology, Innovation \& Ventures, HSBC, 8 Canada Square, E14 5HQ, London, United Kingdom}

\address{~$^{5}$ Quantum Technologies Group, Emerging Technology, Innovation \& Ventures, HSBC, 20 Pasir Panjang Road, 117439, Singapore}

\address{~$^{6}$ Department of Computer Science, Czech Technical University in Prague, Karlovo nam. 13, Prague 2, Czech Republic}

\address{~$^{7}$ Archimedes Research Unit on AI, Data Science and Algorithms, Athena Research Center, 15125, Marousi, Greece}

\address{~$^{8}$ HSBC Global Asset Management, Esplanade Du Gal De Gaulle, Courbevoie Hauts-De-Seine, 92400, Paris, France}

\makeatletter
\renewcommand\paragraph{\@startsection{paragraph}{4}{\z@}%
            {-2.5ex\@plus -1ex \@minus -.25ex}%
            {1.25ex \@plus .25ex}%
            {\normalfont\normalsize\bfseries}}
\makeatother
\begin{document}

\begin{abstract}
We investigate the feasibility of integrating quantum algorithms as subroutines of simulation-based optimisation problems with relevance to and potential applications in mathematical finance.  To this end, we conduct a thorough analysis of all systematic errors arising in the formulation of quantum Monte Carlo integration in order to better understand the resources required to encode various distributions such as a Gaussian, and to evaluate statistical quantities such as the Value-at-Risk (VaR) and Conditional-Value-at-Risk (CVaR) of an asset. Finally, we study the applicability of quantum Monte Carlo integration for fundamental financial use cases in terms of simulation-based optimisations, notably Mean-Conditional-Value-at-Risk (Mean-CVaR) and (risky) Mean-Variance (Mean-Var)  optimisation problems. In particular, we study the Mean-Var  optimisation problem in the presence of noise on a quantum device, and benchmark a quantum error mitigation method that applies to quantum amplitude estimation -- a key subroutine of quantum Monte Carlo integration --  showcasing the utility of such an approach. 
\end{abstract}

\maketitle

{\hypersetup{linkcolor=black}
\setcounter{tocdepth}{1}
\makeatletter
\def\l@subsection{\@tocline{2}{0pt}{2.5pc}{2.5pc}{}}
\makeatother
\tableofcontents
}

\section{Introduction}

Financial institutions routinely face a multitude of problems that involve numerical computations. 
These include forecasting problems, such as pricing, risk estimation, anomaly detection, and understanding customer preferences, as well as optimisation problems such as portfolio selection, optimal trading-strategy development, and hedging. 
More often than not, such problems include significant theoretical and practical challenges that are difficult to overcome. 
These challenges arise from both the complexity associated with the underlying financial models and the limitations of traditional computational methods, even in the current era of advanced parallel computing platforms enabled by modern GPU architectures. 
Investigating how such challenges could be tackled using quantum computing (QC) is currently an overarching effort of the quantum community, and in particular applications of QC for financial modeling are of great promise \cite{egger2020quantum, herman2022survey}.

\subsection*{Quantum computing for financial modeling}
Both academia and industry have made remarkable progress in exploring the potential of QC to address problems that are classically computationally resource intensive or difficult to solve, offering speedups ranging from polynomial to exponential  \cite{Groversearch, Shor,childs2003exponential}. The focus has predominantly been on protocols for optimisation, machine learning, and stochastic simulation, leading to the development of various QC algorithms tailored to financial applications that make use of the above protocols as underlying subroutines. For instance, quantum algorithms for Monte Carlo integration (MCI) promise a quadratic speedup in convergence compared to classical methods, offering significant prospects in risk modelling, where Monte Carlo methods are vital \cite{brassard2000quantum, montanaro2015quantum, Herbert2022,intallura2023survey, Herman_2023}. For optimisation and machine learning, certain quantum algorithms have shown advantages in discrete optimisation, dynamic programming, boosting, and clustering, sometimes demonstrating polynomially-improved complexity compared to classical algorithms \cite{abbas2023quantum}. Furthermore, heuristic approaches, frequently employing classical-quantum hybrid methodologies that make use of variational quantum circuits, have been actively researched \cite{abbas2023quantum, Shaydulin_2024}. Given the broad applicability of these algorithms, great promise is held for financial applications. Within finance, the inherent strengths of QC make it particularly suited for addressing challenges in computational finance with examples such as risk analysis, risk assessment and portfolio management \cite{woerner2019, egger2019credit, stamatopoulos2022towards, matsakos2023quantum}, option and derivative pricing \cite{stamatopoulos2020option, chakrabarti2021threshold, akhalwaya2023modular}, and collateral optimisation \cite{giron2023approaching}. Although achieving fully fault-tolerant quantum computation remains an ongoing challenge, progress is being made \cite{PhysRevX.11.041058, Bluvstein2023, ruiz2024ldpccat, da2024demonstration}, with the aim of running practical applications on quantum machines \cite{preskill2018quantum}. 

While the potential of QC for finance is promising \cite{egger2019credit,mckinseyQuantumTechnology}, it is important to acknowledge that there exist several current limitations and challenges. On the quantum-algorithmic side, solving optimisation problems using quantum computers remains challenging, often requiring the use of heuristic algorithms run on near-term hardware \cite{abbas2023quantum}, which offer no guarantees on convergence times or approximation ratios. Nevertheless, there are continual development of new protocols for quantum optimisation that provide empirical confidence for potential computational advantage; see e.g., Ref.~\cite{montanezbarrera2024universal}. 
In this paper, we focus on the application of QC to problems related to optimisation in financial modeling, and thus we must first introduce the topic and discuss some of its unique challenges.

\subsection*{Optimisation in financial modeling}
Optimisation in financial modeling is of great importance for applications within finance, however it is made difficult by a number of fundamental challenges. One challenge is that data is often high dimensional. For example, in portfolio optimisation and risk management, often one encounters the problem of estimating the covariance matrix (or its inverse) of the asset returns in the portfolio. For instance, with $500$ stocks under consideration for asset allocation, the covariance matrix has $125,250$ parameters \cite{campbell1998econometrics}. Furthermore, phenomena that occur in high-dimensional data, such as collinearities or spurious correlations, add significant complications. This challenge is often referred to as the curse of dimensionality and it is a fundamental limitation, especially in applications within risk management and derivative pricing. Intensive Monte Carlo simulations \cite{Glasserman2003, giles2018multilevel} are the main tool employed in these scenarios, and are considered well suited for tackling this problem -- often facilitated by modern GPU architectures to expedite computations.
evertheless, while GPUs can accelerate such computations, they do not mitigate the fundamental superpolynomial or exponential growth in complexity associated with increasing dimensionality, and thus can at best offer constant factor speedups.  

Another fundamental challenge is that non-convex problems are usually $\textbf{NP}$-hard. For example, the following simple concave  quadratic program
        \begin{maxi*}
        {x \in \R^n}{\frac{1}{2} {x}^\intercal Q {x}+{c}^\intercal {x} }{}{}
        \addConstraint{A  {x} }{\leqslant  {b},}{}
        \end{maxi*}
where $c\in \mathbb{R}^n$, $A\in \mathbb{R}^{m \times n}$, ${b} \in \mathbb{R}^m$  and $Q$ is a $n \times n$ symmetric negative semi-definite matrix, is $\textbf{NP}$-hard \cite{Sahni1974}.
Even if one relaxes the condition on $Q$, such that it is a symmetric matrix with only a single negative eigenvalue, the problem remains $\textbf{NP}$-hard \cite{Pardalos1991}. Such non-convex problems do not provide guarantees that any local minimum is a global minimum, which significantly complicates obtaining globally optimal solutions. 

In financial modeling, a simple example of an $\textbf{NP}$-hard problem is the two-player zero-sum game (related to other models studied later in this paper)
    \begin{mini*}
        {x \in \R^n}{\max _{\mu \in[\hat{\mu}-\delta, \hat{\mu}+\delta]} \left(R^{(f)}-\mu^\intercal  {x}\right)-\frac{\gamma}{2} {x}^\intercal  {\Sigma} {x}}{}{}
        \addConstraint{ }{\mathbf{1}^\intercal {x}}{=1}
        \addConstraint{}{\quad {x} }{\geq 0,}
    \end{mini*}
where $x$ is a vector of  decision variables, $\mu \in \mathbb{R}^n$ (equivalently, $\hat{\mu}$) is the (estimated or expected) mean return vector of the assets under consideration, $\gamma \in \mathbb{R}_{+}$  is the so-called risk-aversion coefficient, $\Sigma \in S_{+}^n$ is the covariance matrix, $\delta \in \mathbb{R}_{+}$ is a scalar that quantifies the uncertainty or allowable deviation in the mean return vector $\mu$ and $R^{(f)}$ is the risk-free return. 
In this problem, one agent seeks to minimise the maximum potential \emph{regret}, defined as the difference between the best possible outcome and the actual outcome, while an adversarial agent aims to maximise this regret by selecting the least favorable expected returns. 
The non-convexity arises due to the max-min structure; while the inner maximisation problem is convex (linear in $\mu$), when combined with the outer minimisation over the binary decision variables, it results in a non-convex problem overall \cite{BenTal1988}. 

An extension of this optimisation problem can be obtained by considering other measures of uncertainty, market fluctuations, and the underlying distributions and their parameters.   
Optimising in the presence of such uncertainties can broadly be formulated as the following optimisation problem:
\begin{mini}
{x \in \X}{f(x) + \mathbb{E}\left[g_0(x, {\xi})\right] }
{}{}
\addConstraint{\mathbb{E}\left[g_i(x, {\xi})\right] \leq 0,}{ \quad i=1,\cdots, k},
\label{eq:sp-model}
\end{mini}
where $x \in \X \subseteq \mathbb{R}^{n}$ is a vector of (possibly discrete) decision variables, $\xi \in \Xi \subseteq \mathbb{R}^m$ a random variable that models the uncertainty of the environment in the problem of interest, $f: \X \to  \mathbb{R}$ a real-valued deterministic function of $x$ and $g_i : \X \times \Xi \to \mathbb{R}$, $i=0,1,\cdots,k$ with $k \in \mathbb{N}$ and $k \le \infty$ is a real-valued stochastic function of $x$ and $\xi$ that captures the uncertainty of the model. Here, $g_0(x,\xi)$ represents the stochastic component in the objective function and $g_i(x,\xi),i=1,\cdots,k$ represent the stochastic components in the constraints which can be independent or correlated.
Assuming that the uncertainty $\xi \sim P$, for some known distribution $P$ that is independent of $x$, then the stochastic term $g_0$ could be, for example, the value-at-risk (VaR), conditional value-at-risk ($\cValueAtRisk$) or some high-order moments of $\xi$. In practice, these optimisation problems correspond to a wide range of risk measure calculations and portfolio optimisations. For example, ``Maslowian Portfolio Theory'' argues that a coherent risk measure such as $\cValueAtRisk$ can be used for the optimisation of investment portfolios, see Refs.~\cite{debrouwer2009maslowian,debrouwer2012maslowian,debrouwer2016proposal}.

The general framework for such optimisation problems falls under the umbrella of stochastic programming  (SP) \cite{Nemirovski2007,Shapiro2021-ay}. SP is a well-established approach specifically designed to address optimisation problems under such conditions. Unlike deterministic optimisation, SP is able to take into account inherent uncertainty in financial markets, and thus seeks solutions that are robust for a range of possible scenarios \cite{shao2001stock}. It turns out that there exists a family of SP problems for optimisation that are particularly amenable to be solved using QC, that of simulation-based optimisation (SBO).

\subsection*{Quantum computing for simulation-based optimisation }
SBO, in which stochastic approximations are employed to evaluate functions of random variables within the context of an otherwise deterministic optimisation problem, can be seen as the set of SP problems for which the computation of the terms that involve random variables cannot be performed analytically.  The goal is to simulate a sequence of system configurations, each corresponding to specific settings of the decision variables, in order to identify an optimal or near-optimal configuration. This approach allows for the evaluation of a small subset of possible configurations, thereby avoiding the computational infeasibility of exhaustive enumeration \cite{Law2002simulation}. This is therefore useful for families of optimisation problems that involve scenarios where in the objective function -- and potentially in the constraints -- one encounters quantities that depend on moments of random variables \cite{Gosavi2015,homem2014monte}. Reference~\cite{gacon2020quantum} first considered how quantum amplitude estimation (QAE) and quantum Monte Carlo integration (QMCI) can be used to accelerate SBO over both continuous and discrete variables, introducing a quantum-enhanced SBO algorithm. Concretely, the problem of interest therein is essentially similar to  that of \eqref{eq:sp-model}, but assuming no access to the analytical form of $\mathbb{E}[g_i(\xi)]$. 
 The authors of Ref.~\cite{gacon2020quantum}\footnote{Comparing the notation of \cite{gacon2020quantum} to ours the equivalence is $y \to x$ and $X \to \xi$. Their formulation also does include deterministic terms on the objective or stochastic terms as constraints.} discriminate between two approaches depending on the nature of the search variable $x$; continuous or discrete. In the former case they use modern variations of the QAE algorithm; see Section~\ref{sec:quantum_alternatives} for a brief exposition of the ones relevant to our work. In the latter, where the search variable $x$ is discrete, they solve the optimisation part of the problem by means of a parametrised quantum circuit (PQC).

The interest in SBO arises due to its wide range of applications in asset management, including optimal asset allocation, risk-return trade-offs, asset-liability management (e.g., cash flow matching), portfolio optimisation (e.g., robust portfolio construction and risk management), stress testing, risk assessment, and scenario analysis (scenario-based optimisation); see \cite{Gosavi2003,Adachi2005,Bao2014}. Given the difficulty of the problem, it is only natural to ask whether a practical realisation of quantum-enabled SBO for these applications is possible, and answering this question is the primary motivation for this paper.

\subsection*{Our contribution}
Going a step beyond what is considered ``usual'' in the quantum finance literature, our work aims to investigate how QMCI can assist optimisation problems that involve elements of risk factors; that is, include more a more realistic approach to optimisation problems that appear in financial modelling. To use QMCI for practical applications it is essential to understand all the sources of systematic error that are introduced at each stage of the workflow. 
This ensures that the final results are interpretable and that the precise bottlenecks for a particular use case are understood. To the best of our knowledge, a thorough analysis of all systematic errors introduced when running QMCI has not previously been carried out -- we refer to \cite{chakrabarti2020threshold} for previous initial studies in this direction. In addition, for near-term applications of QMCI to optimisation, then given the limitations of current hardware the negative effects of noise will be significant; this can be considered as a significant future ``bottleneck'' to successfully deploying QMCI for optimisation in the near term. Thus exploring methods of quantum error mitigation that can reduce the impact of noise will be of key importance.

To this end, we explore the strengths and limitations of quantum-enhanced SBO for some well-known financial modeling problems that are often associated with quantum finance, namely the family of ``risky'' portfolio-optimisation problems. This contributes to the ongoing exploration of how to integrate QC into financial modelling, paving the way for further advancements for applications in portfolio construction. 
Our main contributions are as follows:

\begin{enumerate}
\item We identify and define the systematic errors involved in QMCI, which can easily be extended to state preparation of any quantum computation, then we provide detailed analyses of the systematic errors to provide insights in to the performance of QMCI.
The systematic errors are both related to the distribution that is integrated over and also to the operation of the algorithm that is the integration over a specified region. 
To this end, we decompose the quantity that is \emph{actually} being estimated into \emph{intermediate} quantities, which are helpful when we want to write down and calculate the general error definitions for a specific function that is integrated.   
\item We examine the feasibility of using QMCI under resource constraints -- particularly limited qubit counts -- and apply our systematic error analyses to different use cases revealing further insights into the performance of QMCI and its practical bottlenecks.
In particular, we investigate (C)VaR estimations and the related detailed analyses of the specific systematic errors that arise in this calculation; these not only fully explain the observed numerical results -- functions of the various bottlenecks that exist for the calculations -- but also suggest how these bottlenecks can be overcome. 
Our key conclusion is that QMCI provides accurate estimations and exhibits the expected quadratic advantage in terms of error scaling as compared to CMCI, but realising this advantage in practice is contingent on suppressing systematic errors to a sufficient degree.
\item We investigate practical SBO problems using QMCI, in which  we make use of QMCI as a subroutine for the SBO. Given the relative lack of detailed SBO studies that exist in the literature, following \cite{gacon2020quantum} we consider two different use cases concerning asset allocation under uncertainty and provide numerical results for each use case. We start by solving a mean-CVaR portfolio optimisation problem, showcasing the effectiveness of QMCI as compared to classical Monte Carlo integration (CMCI). We then solve a Mean-Var optimisation problem for asset allocation, in which we benchmark results in the presence of device noise using a method of quantum error mitigation that applies to the QAE subroutine of QMCI,  demonstrating a more realistic example of how one might actually perform SBO on near-term quantum hardware.\footnote{Indeed, we can consider the effects of noise as an additional type of error for near-term QMCI calculations.} This is also the first time that this method of error mitigation has been applied to a real-world problem that makes use of QMCI as a subroutine. Based on this study we conclude that our method of error mitigation for QAE does improve the expected results when considering estimation of the objective function. However, when considering the full SBO problem -- which involves both optimisation and estimation -- then we cannot determine conclusively if there is any improvement, given the limited statistics available for the study and the additional complexity of the full SBO problem, where error contributions from sources both related and unrelated to noise are compounded.
\end{enumerate}

It is worth noting that our work in general is significantly differentiated from that of \cite{gacon2020quantum}. For example, in terms of portfolio optimisation \cite{gacon2020quantum} studied a $\ValueAtRisk$-based binary optimisation problem for asset allocation, while in our work we focus on the estimation of $\ValueAtRisk$ and $\cValueAtRisk$ over a number of distributions.

\subsection*{Organisation}In order to provide the right level of context and mathematical tools, we introduce a glossary of financial concepts in Section~\ref{sec:glossary-mathematical-finance} and we introduce the relevant quantum tools, notably QMCI, used throughout this work in Section~\ref{sec:quantum_alternatives}. Subsequently, we provide a detailed description of the systematic errors involved in the methodology of QMCI in Section~\ref{sec:systematic-errors-section} followed by (C)VaR estimations in Section~\ref{sec:var-cvar-estimations} as well as the related systematic error analyses.
We then investigate two practical SBO problems using QMCI to solve a mean-CVaR portfolio optimisation problem in Section~\ref{sec:mean-cvar-SBO} and then to solve a Mean-Var optimisation problem for asset allocation in the presence of device noise in Section~\ref{sec:asset-allocation-uncertainty}. 
We conclude the paper in Section \ref{sec:conclusions}.
 


\section{Glossary and definitions of financial concepts} \label{sec:glossary-mathematical-finance}
To bridge the gap between the mathematical finance literature and the quantum computing literature, we present in this section a glossary containing relevant definitions.

We denote the general variables:
\begin{enumerate}
    \item The return of a risk-free asset or portfolio $R^{(f)}$. A risk-free asset or portfolio is a theoretical financial instrument that is assumed to have no risk of financial loss and is expected to return its full value with interest at maturity. When the risk-free return is considered over different periods, we add an subscript $R_T^{(f)}$ to refer to the time period $T$.
    \item The return of a portfolio $X$ which represents the gain or loss on the portfolio over a specific period, often expressed as a percentage. This return can be calculated based on the weighted average of the returns of the individual assets within the portfolio, where the weights correspond to the proportion of each asset's value relative to the total value of the portfolio.
    \item  The return of a portfolio admits different decompositions, a possible one is $X=r+R^{(f)}$ where $r$ is the risk premium or excess return of the portfolio - it describes the fluctuations induced by the risk. 
    \item For a portfolio of a single asset, the value/wealth of the asset at time $T$ is denoted as $V_T$.
    \item The return of the risky asset during the period $T$ is denoted as $X_T$. 
    \item Logarithmic returns, also known as log-returns, which measures the percentage change in the value of an asset over a period of time.  The log-returns can be calculated by $X_{T+1} = \log\left(\frac{V_{T+1}}{V_T}\right)$.
    \item For a portfolio of multiple assets, the return of asset $i$ is denoted as $X_i$.
    \item The utility function $U$ over which we optimise.\\
    \end{enumerate}

The following probabilistic quantities have the following meaning in the context of this work:
    \begin{enumerate}
    \setcounter{enumi}{8}
    \item $\E[X]$ the expectation of the value of the return/wealth of an asset.
    \item $\ValueAtRisk$ the Value-at-Risk of a portfolio. For a random variable $X$, $ \ValueAtRisk_{\alpha}(X) := \inf \{\gamma : \Pr(X \le \gamma) \ge \alpha \}$ for the confidence level $\alpha \in (0,1)$. $\ValueAtRisk_\alpha(X)$ is a lower percentile of the random variable $X$ for a given loss distribution.
    \item $\cValueAtRisk$ is the Conditional-Value-at-Risk which quantifies the expected loss exceeding the Value-at-Risk. For a random variable $X$, $\cValueAtRisk_{\alpha} (X) :=\E[X|X \ge \ValueAtRisk_{\alpha}(X)]$ for the confidence level $\alpha \in (0,1)$.
    \item $\mathcal{N}(\mu,\sigma^2)$ a Gaussian distribution with mean $\mu$ and variance $\sigma^2$ with distribution $f(x)=\frac{1}{\sqrt{2\pi}\sigma}e^{-\frac{(x-\mu)^2}{2\sigma^2}}$.
    \item $\mathcal{LN}(\mu,\sigma^2)$ a log-normal distribution with parameters $\mu$ and $\sigma^2$, whose probability density function (PDF) is $f(x)=\frac{1}{x \sigma \sqrt{2\pi}\sigma}e^{-\frac{(\ln{x}-\mu)^2}{\sigma^2}}$.
    \item $\mathcal{L}(\mu,c)$ a L\'evy distribution with location parameter $\mu$ and scale parameter $c$ with distribution $\mathcal{L}(x,\mu,c)=\sqrt{\frac{c}{2 \pi}} \frac{e^{-\frac{c}{2(x-\mu)}}}{(x-\mu)^{3 / 2}}$, for $x \geq \mu$.\\
    \item $\Theta(y)$  the indicator function defined as
\begin{equation}
    \Theta(y) = \begin{cases}
        1, & \text{if}~ y \ge 0 \\
        0, & \text{otherwise},
    \end{cases} \label{eq:indicator_var}
\end{equation}
where $y$ can be either a variable or an expression like $y = x - x_{th}$ with a constant $x_{th}$. In this sense, $\Theta(y)$ can be equivalently represented as $\Theta(x - x_{th})$ or $\Theta(x \ge x_{th})$ for $y = x - x_{th}$. 
    \end{enumerate}

In the case of asset allocation under uncertainty:
    \begin{enumerate}
    \setcounter{enumi}{15}
    \item $w=\{w_i \}_{i=1}^n $ an allocation weight over $n$ assets. If no leverage is allowed in the allocation (no short position can be taken), then $w_i\geq 0$ for all $i\in [1,n]$ and $\sum_{i=1}^{n} {w_i} =1$.
    \item The return of a portfolio with a specific allocation $\sum_{i=1}^n w_i X_i$.
    \item $\lambda \in \R$ a Lagrange parameter describing the level of risk aversion of an investor.
\end{enumerate}

\section{Quantum enhancements of Monte Carlo integration}\label{sec:quantum_alternatives}

\subsection{Monte Carlo integration}

The task in MCI is to estimate the value of an integral over a measured space $\Omega$, often a subspace of $\R^n$. 
Assuming we are interested in integrals over the real line, the expectation of a function $g(x)$ of a continuous random variable $X$ with the PDF $f_{X}(x)$ is defined as
\begin{equation}\label{Eq:FunctionAppliedIntergral}
    \mathbb{E}\left[g(X)\right] = \int_{-\infty}^{\infty}g(x)f_{X}(x)\,dx.
\end{equation}
We refer to $g(x)$ as the ``function applied''.
On any digital computer, the support of the distribution must be discretised and truncated, and the functions may only take discrete values.\footnote{This applies to both the classical and quantum cases, but because resource management is more involved in QC, more careful consideration must be taken, and this is discussed in detail in this work.}
This means that both classical and quantum digital computers can only approximately prepare the distribution and compute the function applied.
Therefore, MCI on a digital computer estimates $\mathbb{E}\left[g(X)\right]_{S}$, where the subscript $S$ means that the expectation of the function applied is computed for a digitised distribution, and the difference between $\mathbb{E}\left[g(X)\right]$ and $\mathbb{E}\left[g(X)\right]_{S}$ is the systematic error introduced by the digitisation, which is discussed in detail in this work.
Classically, we sample the random variable $\mathcal{S}$ times, calculate the function $g(\cdot)$ for each sample, and obtain an estimate of the integral by averaging these values. 
Denoting the estimated value with $\widehat{\mathbb{E}[g(X)]_{S}}$, we then define the estimation error (which includes statistical and systematic errors) as
\begin{equation}
    \epsilon_{E}(\widehat{g(X)}) = \abs{
        \mathbb{E}\left[g(X)\right] - \mathbb{E}[\widehat{g(X)}]_{S}
    },
\end{equation}
and the root mean-squared error (RMSE) can be defined as
\begin{equation}
\label{eq:RMSE-def}
  \RMSE(\widehat{g(X)}) = \sqrt{\mathbb{E}\left[ \left(g(X) - \mathbb{E}[\widehat{g(X)}]_{S}\right)^2\right])}.
\end{equation}
Classically, the central limit theorem can be invoked to obtain the asymptotic decay of the RMSE  as $\mathcal{O}(1/\sqrt{\mathcal{S}})$, where $\mathcal{S}$ is the number of samples. The quantum counterpart of MCI, i.e. QMCI, provides a quadratic speed up to this convergence, meaning the RMSE decays as $\mathcal{O}(1/\mathcal{S})$.

QMCI is a three step process as described in \cite{akhalwaya2023modular}, and consists of preparing a quantum state that has the values of the probability distribution encoded in its amplitudes, applying the observable function $g(\cdot)$ to each state coherently and then measuring the amplitude of an ancilla qubit, providing an estimate of the integral \eqref{Eq:FunctionAppliedIntergral}. 
Mathematically, the starting point is a circuit, $P$, acting on $n$ qubits, that prepares a quantum state, $\ket{p}$, encoding the probability distribution $f_X(x)$. 
That is, $\ket{p} = P \ket{0}_n$ such that
\begin{equation}
\ket{p} = \sum_{x} \sqrt{f_X(x)} \ket{x}.
\end{equation}
The coherent application of the observable function is provided via a circuit $R$ such that
\begin{equation}
R \ket{p} \! \ket{0} = \sum_{x } \sqrt{f_X(x)} \ket{x}  \left( \sqrt{1-g(x)} \ket{0} + \sqrt{g(x)} \ket{1} \right).
\end{equation}
In Quantinuum's QMCI engine~\cite{akhalwaya2023modular}, the circuit $R$ is implemented using the Fourier QMCI approach~\cite{Herbert2022}.
Finally, the probability of measuring $1$ on the final qubit is $\sum_{x } g(x)  f_X(x)$. Notice that there will be no speedup if we directly measure the ancilla of the state $R \ket{p} \! \ket{0}$, and the number of samples required will be the same as for classical Monte Carlo methods.
QAE is the algorithm used to measure this probability, and it is the core subroutine that provides a quadratic improvement for the rate of convergence of QMCI \cite{akhalwaya2023modular} over CMCI.

\subsection{Quantum amplitude estimation }\label{subsec:qae}
QAE, originally introduced in \cite{brassard2000quantum}, is the key quantum algorithm used for QMCI. Variations of this algorithm have been proposed over time and we present in this Section a brief and standard presentation of it, for consistency. If we consider an unitary operator $\mathcal{A}$ acting on a register of $(n+1)$ qubits $\ket{0}_n\ket{0} \equiv \ket{0}_{n+1}$ such that the following state is prepared 
\begin{align}
\mathcal{A}\ket{0}_{n+1}= \sqrt{1-a}\ket{\psi_0}_n\ket{0} +\sqrt{a}\ket{\psi_1}_n\ket{1},
\end{align}
where the $\ket{\psi_0}_n$ and $\ket{\psi_1}_n$ states are normalised, then QAE can estimate the parameter $a = \sin^2 \theta_a$, $\theta_a \in (0,\pi/2]$ with the convergence of the root mean-squared error (RMSE) of the estimate scaling as $\mathcal{O}(1 / \mathcal{S})$, where $\mathcal{S}$ is the number of quantum samples i.e., the total number of uses or calls to the oracle $\mathcal{A}$. This provides a quadratic advantage as compared to classical Monte Carlo sampling -- the source of quadratic advantage in QMCI as previously discussed.  The algorithm makes use of the Grover operator, first introduced in \cite{brassard2000quantum} 
\begin{align}
    Q=\mathcal{A} \left(\mathbf{1}-2\ket{0}_{n+1} \bra{0}_{n+1} \right)\mathcal{A}^\dagger \left(\mathbf{1}-2 \ket{\psi_0}_n\ket{0} \bra{\psi}_n \bra{0} \right),
    \end{align}
where $\mathbf{1}$ denotes the identity operator. The canonical method for QAE relies on quantum phase estimation (QPE) (which relies on the inverse quantum Fourier transform) to estimate the parameter $a$. This requires significant QC resources -- a substantial number of qubits and controlled-not (CNOT) gates. For near-term intermediate-scale quantum (NISQ) devices this is problematic, and several approaches have been developed to overcome this issue (see \cite{intallura2023survey} for further discussion).\\

One such approach is the maximum-likelihood QAE (MLE-QAE) of \cite{suzuki2020amplitude}, refined by \cite{callison2022improved}. There QPE is no longer required, and is replaced by a classical maximum-likelihood-estimation (MLE) subroutine (see also \cite{aaronson2020quantum} for another alternative without QPE). One prepares
\begin{equation}
Q^{m_k}( \cos \theta_a \ket{\psi_0}_n\ket{0} + \sin \theta_a \ket{\psi_1}_n\ket{1}) = \cos ((2m_k+1)\theta_a) \ket{\psi_0}_n\ket{0} + \sin((2m_k+1)\theta_a) \ket{\psi_1}_n\ket{1},
\end{equation}
for different $m_k = \{m_0,m_2,\ldots m_{M-1}\}$, for $N_k$ measurements (shots) per $m_k$. The set $\{ (m_k, N_k) \}_k$ is usually referred to as the schedule of QAE, and there are several ways of choosing this schedule; for example one schedule introduced in \cite{suzuki2020amplitude} that has near-optimal quadratic convergence is the so-called exponentially increasing sequence (EIS) $\{ (m_k, N_{\text{shots}}) \mid m_k = 2^k, \ k \in \mathbb{N} \}$, for some fixed number of shots $N_{\text{shots}}$, and this is the schedule that is used throughout this paper. Then letting $h_k$ be the number of times the $\ket{\psi_1}_n\ket{1}$ state is measured -- with corresponding probability $\sin^2((2m_k+1)\theta_a)$ -- the likelihood function is
\begin{equation}\label{eq:likelihood}
L_k\left(h_k, \theta_a\right)=\left[\sin ^2\left(\left(2 m_k+1\right) \theta_a\right)\right]^{h_k}\left[\cos ^2\left(\left(2 m_k+1\right) \theta_a\right)\right]^{N_k-h_k}.
\end{equation}
The combination of the likelihood functions $L_k\left(h_k ; \theta_a\right)$  provides a single likelihood function 
\begin{equation}
L\left(h, \theta_a\right)=\prod_{k=0}^{M-1} L_k\left(h_k ; \theta_a\right),
\end{equation}
with $h=\left(h_0, h_1, \cdots, h_{M-1}\right)$. The MLE estimate of $\theta_a$ is then obtained as
\begin{equation}
\hat{\theta}_a=\underset{\theta_a}{\arg \max }\, \log L\left(h ; \theta_a\right).
\end{equation} A slight modification to this approach is to instead set a uniform prior distribution $p(\theta_a)$ across a uniformly-spaced grid of $n$ $\theta_a$ values, and then use Bayesian inference with the following update rule to the posterior distribution
\begin{equation}
p(\theta_a) \leftarrow p(\theta_a) L_{k}(h_k|\theta_a),
\end{equation} (followed by renormalising) to provide an estimate
\begin{equation}
\hat{\theta}_a= \sum^{n}_{i=1} \theta_a^i p^i.
\end{equation} It is such an approach that we make use of in the rest of this paper.

Two additional approaches that do not make use of QPE are also worth mentioning for the purposes of this paper. The approach of iterative QAE of \cite{Grinko_2021} gives a robust theoretical guarantee in terms of convergence. In addition, Ref.~\cite{akhalwaya2023modular} introduced the linear-combination-of-unitary-operators QAE (LCU-QAE); specifically, \cite[Section~5]{akhalwaya2023modular} found evidence that LCU-QAE performs competitively when considering convergence, while also having robust statistical properties of the estimator -- being close to asymptotically unbiased and exhibiting near-Gaussian kurtosis and skewness. QAE still remains an active topic of research and new methods have recently been proposed, for example see \cite{labib2024quantum} for a connection between QAE and classical signal processing.

\subsubsection{Noise-aware quantum amplitude estimation}
\label{subsec:naqae}
During the NISQ era, the prominent characteristic of quantum computation is its susceptibility to noise and errors. A key measure quantifying to what extent noise affects operations, and thus how badly noise can affect a given computation, is the two-qubit gate infidelity, $1-\varphi_2$, where $\varphi_2$ is the two-qubit average gate fidelity, defining the average degree of accuracy between the ideal operation of a two-qubit gate and its actual implementation on real quantum hardware. The infidelity thus quantifies the degree of \textit{deviation} between the ideal operation and its actual implementation. 

Current NISQ devices have high error rates (i.e., large gate infidelities). To address the challenge of noise affecting computations, quantum error correction protocols have been proposed. However, implementing fully-fledged quantum error-correction (QEC) codes requires a considerable overhead in qubit resources and computational complexity, something not currently feasible on NISQ-era devices due to their limited qubit count and high error rates.

Error mitigation techniques thus have a role in improving the reliability of quantum computations without necessitating full QEC. Error mitigation strategies aim at reducing the impact of noise and errors on the outcomes of quantum algorithms run on noisy devices. Unlike QEC, which demands significant overhead and resources, error mitigation techniques offer intermediate solutions that are more feasible for current NISQ-era devices. A well known example is zero-noise extrapolation \cite{Giurgica_Tiron_2020}, which relies on post-processing the outcomes of a noisy computation in order to mitigate the impact of the noise on the final results.

Reference~\cite{herbert2021noise} proposes a means of error mitigation that can be applied to any QAE algorithm that uses repeated shots of the same circuit to infer the amplitude, such as the MLE-QAE introduced in the previous section (we will now only consider MLE-QAE in this example). The approach is termed noise-aware QAE (NA-QAE), and involves directly translating the noise at the physical level to the resultant estimation uncertainty at the application level, such that the QAE algorithm is ``noise aware''. 

In this approach the noise introduced by applying the Grover operator $Q$ in QAE is modelled as Gaussian noise affecting the measured qubit, parameterised by two constants $k_\sigma$ and $k_\mu$, which encode the properties of the noise, such that that after $m$, applications of the Grover operator the following state is prepared \begin{align}
    Q^{m_k}&( \cos \theta_a \ket{\psi_0}_n\ket{0} + \sin \theta_a \ket{\psi_1}_n\ket{1}) = \nonumber \\ &\text{cos}((2m_k+1)\theta_a + \theta_{a}^{\epsilon})\ket{\psi_0}_n\ket{0} +\text{sin}((2m_k+1)\theta_a + \theta^{\epsilon}_{a})\ket{\psi_1}_n\ket{1},
\end{align}
where $\theta_a$ corresponds to the desired rotation corresponding to the unitary enacted by $Q$ and $\theta_{\epsilon} \sim \mathcal{N}(k_\mu m_k ,k_\sigma m_k)$ is an additional erroneous component to the rotation arising from the noise. The probability of measuring the circuit in the $\ket{1}$ state is then modified to $e^{-2k_{\sigma} m_k}\sin^2((2m_k+1)\theta_a + k_\mu m_k)$ and the likelihood modified accordingly.

If an assumption is made that $k_{\mu}=0$ (by previous calibration of the hardware for example), then the error introduced by the noise only arises from the ``variance'' component proportional to $k_\sigma$. This component can be suppressed by increasing the number of shots run. Thus NA-QAE proposes a methodology for the factor increase in the number of shots required to obtain the same estimation worst-case variance for the estimate of the amplitude as one would have achieved in the noiseless case when running $N_{k}$ shots
\begin{equation}
N^{\text{na-qae}}_{k} = (4k_\sigma m_k+1)N_k.
\end{equation}

Of course, implementing this update rule requires knowledge of the parameter $k_\sigma$, and this may be possible to calculate from device characteristics alone, under certain assumptions. If the assumption $k_{\mu}=0$ is made, then the Gaussian noise model is equivalent to depolarising noise affecting a single qubit. In this case one can equate $k_{\sigma} = - \text{ln}(\tilde{p}_{\text{coh}}) / 2$, where $\tilde{p}_{\text{coh}}$ is the probability that the state remains coherent after each application of $Q$ under a depolarising noise model. Then (ignoring 1-qubit gates and other sources of error such as memory errors and state preparation and measurement (SPAM) errors) it can be shown that the coherence probability per application of $Q$ can be approximately expressed as \cite{Sanders_2016}
\begin{equation}\label{eq:k_sigma_dep}
\tilde{p}_{\text{coh}} =  [1 - 2(1-\varphi_{2})]^{n_{2}},
\end{equation}
where $n_{2}$ is the total number of two-qubit gates per application of $Q$. This then gives a method to estimate $k_\sigma$ from the device characteristics. However, in the approach of MLE-QAE, all that is required is that a prior distribution is placed on $k_\sigma$ in order to construct a joint prior distribution $p(\theta_a, k_\sigma)$, and then as the algorithm progresses by running batches of circuits for different $m_k$ values, $k_\sigma$ can be continuously inferred and updated. In this way, the algorithm is adaptive because, as the noise profile is updated based on the data for a given batch of circuits, the corresponding factor increase in the number of shots to run for the next batch of circuits is then inferred, and so on. 

Because the performance of QAE is key to the performance of QMCI, techniques such as NA-QAE will be very useful in the NISQ era when it comes to being able to run real-world problems on quantum hardware. Later on in Section~\ref{sec:asset-allocation-uncertainty}, we will demonstrate the first application of NA-QAE for performing QMCI for a problem of real-world interest, based on a simplified SBO problem for portfolio optimisation for asset allocation under uncertainty.

\subsection{Quantinuum's quantum Monte Carlo integration engine}

In this work we use the end-to-end engine developed and originally introduced in Ref.~\cite{akhalwaya2023modular} to investigate practical applications of QMCI. 
This section provides a high level overview of the engine and we refer to Ref.~\cite{akhalwaya2023modular} for further detail. The engine was designed to provide a modular tool to perform QMCI calculations. Its modularity comes from its division into multiple modules, represented in Ref.~\cite[Fig.~1]{akhalwaya2023modular}, which can be easily updated or replaced, independently of the other modules. There are two notable advantages to this approach: (1) a new bespoke method for e.g., state preparation can be studied and rapidly prototyped and benchmarked against other methods, and (2) given continuous improvements in the field for QAE methods, noise mitigation or state preparation, we can continuously revisit previously investigated use-cases and see how new methods impact their solution. These advantages are further interesting as the QMCI engine is endowed with a resource quantification mode which accounts for all resources (including qubit and gate counts) required to produce a certain calculation up to a given precision. Hence, the modularity allows us to determine the best method to approach a particular QMCI calculation. The engine provides a library of state-preparation methods, functions applied and QAE methods to facilitate the investigation of applications for which QMCI is either a subroutine (such as in a hybrid classical-quantum optimisation problem) or the expected output (as in the pricing of a financial instrument).

\subsubsection{State preparation with parameterised quantum circuits} 
\label{subsec:param-quantum-circuit}

A critical component of the QMCI engine is its library of state-preparation methods that encode probability distributions in the amplitudes of qubit states. 
As more applications will make use of QMCI over time, the range of probability distributions of interest will also increase. 
For this reason, the library of methods to load probability distributions into quantum states should be as general as possible, both in terms of the types of distributions and the quantum routines used. 
There are several approaches to perform state preparation, we refer to Refs.~\cite{plesch2011quantum, Sanders_2019, zoufal2019quantum,zhang2022quantum,bausch2022fast,rattew2022preparing,wodecki2024spectral,BenedettiRosenkranzStatePrep2024} and references therein for a broad summary of the state of the art.

As regards the QMCI engine, Ref.~\cite{akhalwaya2023modular} describes various distribution-loading techniques, including efficient circuits to load Gaussian and log-normal distributions obtained by training PQCs. 
Ref.~\cite{akhalwaya2023modular} considers two types of ansatz for PQC training: (i) Hardware-efficient (HWE) ansatz and (ii) Tensor network (TN) methods with brickwall ansatz, and we use both methods in this work.

In the next Section, we proceed to a full audit of the errors introduced by the distribution loading process.
In particular, we demonstrate that it is key to understand all state-preparation errors that arise.
The state-preparation (or equivalently distribution-loading) step of QMCI is still the subject of active research in the quantum community, as it is currently the bottleneck for extending to higher-dimensional problems or unstructured distributions. 
Recently, there have been numerous attempts to tackle this problem using more quantum-native approaches such as solving partial differential equations \cite{akhalwaya2023modular}, determining the ground state of supersymmetric Hamiltonians \cite{mazzola2021sampling,MarkJackson2024}, or using the techniques of linear combination of unitaries \cite{BenedettiRosenkranzStatePrep2024}.  
We expect that any improvement as regards state-preparation methods will greatly expand the applicability of QMCI.

\section{Systematic errors in quantum Monte Carlo integration}
\label{sec:systematic-errors-section}

\begin{figure}[t!]
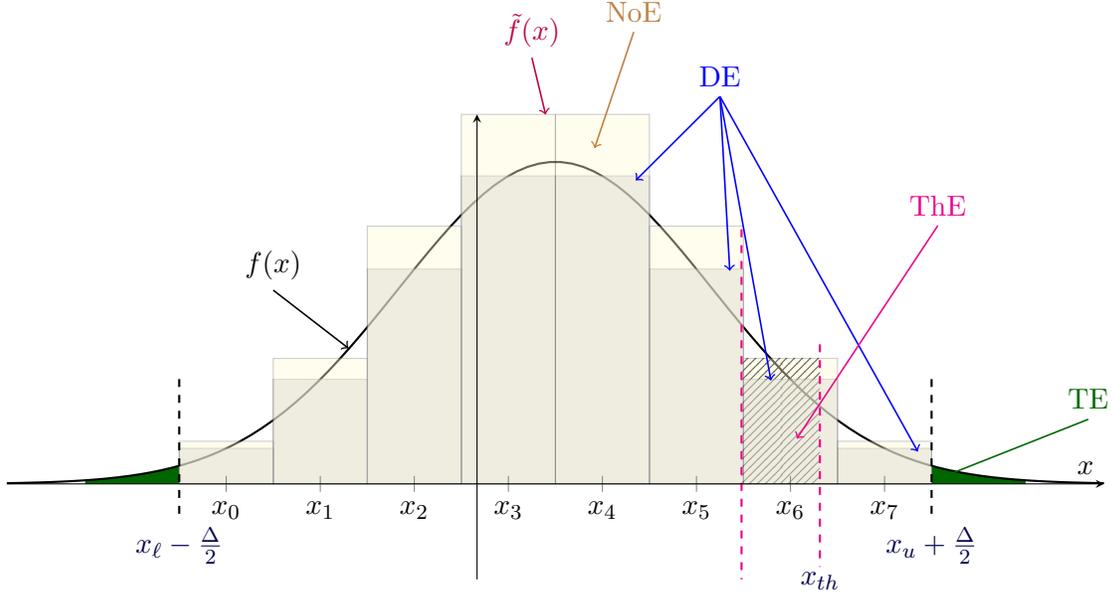

\centering
\errorgraph
\caption{Illustration of the systematic errors in QMCI for a probability density $f_{X}(x)$ over the domain $x\in [-\infty, \infty]$, which includes four components: truncation error (TE), discretisation error (DE), normalisation error (NoE) and thresholding error (ThE). 
TE arises when $f(x)$ is prepared in the limited range $[x_\ell - \Delta/2,x_u + \Delta/2]$, which corresponds to the green area. 
DE arises from the discretisation of $f(x)$ over $N = 2^n$ supports using $n$ qubits over the range, which is the difference between the area covered by $f(x)$ and the area covered by the grey rectangles under the $f(x)$ curve.
As illustrated by the yellow area, NoE arises from the renormalisation of the truncated and discretised function $f(x)$, where $\tilde{f}(x)$ represents the renormalised $f(x)$.  
ThE is not an intrinsic error associated with the representation of the probability distribution; but rather a further ``downstream'' error that will be incurred if a threshold is taken at a value that doesn't exactly align with the histogram interval edge points.} \label{fig:systematic_errors}
\end{figure}

In this section, we define the fundamental systematic errors that arise when we want to estimate the value of an integral using QMCI. Our methodology here is to first define each error in terms of its source at a high level, and then define a measure that quantifies it in the context of QMCI. Different errors with their associated sources are depicted in Fig.~\ref{fig:systematic_errors}.
It is important to note that some of these errors occur both in quantum and classical MCI. 
Also, the errors are not necessarily independent of each other; indeed, the majority of the errors \emph{are} dependent, and while it may be possible to define errors such that they are independent, our measures are designed in order to aid practical use, leading to intuitive measures that are interpretable and thus can provide further insights into results of QMCI.
In each error definition, we decompose the quantity that is \emph{actually} being estimated into \emph{intermediate} quantities that help us to intuitively understand and interpret the errors. 
The \emph{intermediate} quantities are helpful when we want to write down and calculate the general error definitions for a specific function applied. In Sections~\ref{sec:var-cvar-estimations} and Appendix~\ref{app:levy_distribution_section}, we start from an \emph{intermediate} quantity to both calculate the errors for a specific function applied and write an error tailored for the specific application, demonstrating both the importance of the \emph{intermediate} quantities and the adaptability of our approach to practical requirements of different use cases.

\subsection{Truncation error} 
The first error occurs when we truncate the limits of the estimated integral.
For example, one or both integral limits may not be finite (as in \eqref{Eq:FunctionAppliedIntergral}), and truncation of the integral limits to a finite range leads to this error.
If we restrict the limits of the integral in \eqref{Eq:FunctionAppliedIntergral} to the range $\left[a, b\right]$ (and if $f_{X}(x) \neq 0$ for any $x \not\in \left[a, b\right]$), then this error is defined as
\begin{equation}
    \epsilon_{TE}(g(X)) = \abs{\int_{-\infty}^{\infty}g(x)f_{X}(x)\,dx - \int_{a}^{b}g(x)f_{X}(x)\,dx}.
\end{equation}
Now, the intermediate quantity that we say we want to estimate is the following integral
\begin{equation}\label{Eq:TrucantedExpectationGeneral}
    \mathbb{E}\left[g(X)\right]_{T} := \int_{a}^{b}g(x)f_{X}(x)\,dx.
\end{equation}
The truncation error is represented by the green filled areas at the tails of the distribution in Fig.~\ref{fig:systematic_errors}.

\subsection{Discretisation error} 
In MCI, we obtain an estimate of the integral by sampling the random variable $X$. However, the sample resolution might be limited by the finite number of bits used, leading to a discretisation error.
In other words, this error arises when we can only sample the continuous random variable discretely due to the finite number of bits/qubits that encode the random variable.
On a classical computer, we have access to (effectively) an unlimited number of bits and can easily suppress this error to reach a desired accuracy in the estimate.
However, the same cannot be said for a quantum computer, and this error plays a crucial role in QMCI due to the limited number of qubits used to represent the random variable.

For the integral in \eqref{Eq:FunctionAppliedIntergral}, we define the discretisation error as
\begin{equation}\label{Eq:integrationToRiemannSum}
    \epsilon_{DE, n}(g(X)) = \abs{\int_{a}^{b}g(x)f_{X}(x)\,dx - \sum_{i=0}^{N-1}g(x_{i})f_{X}(x_{i})\Delta},
\end{equation}
where $x_{0} := a + \cfrac{\Delta}{2}$, $x_{i} := x_{0} + i\times\Delta$ (such that $x_{N - 1} := b - \Delta/2$), $\Delta := x_{i+1} - x_{i} = \cfrac{b-a}{N}$, and $N = 2^{n}$ for $n$ qubits.
With these definitions, the discrete sum
\begin{equation}\label{Eq:discretized_expectation}
    \mathbb{E}\left[g(X)\right]_{D, n} := \sum_{i=0}^{N-1}g(x_{i})f_{X}(x_{i})\Delta,
\end{equation}
is the next intermediate quantity, which is the mid-point Riemann sum approximation of the integral in \eqref{Eq:TrucantedExpectationGeneral}.\footnote{The mid-point Riemann sum is one of the most common techniques to approximate a definite integral by a finite sum using rectangular regions due to its simplicity and accuracy. The choice of mid-point Riemann sum for the discretisation is unrelated to the QMCI engine, which is agnostic to this choice, and a different choice may be used if it is more suitable for a given function applied.}  The subscript $D$ here refers to ``discretisation''. In the QMCI engine and in the rest of this work, we denote $x_{\ell} = x_{0}$ and $x_{u} = x_{N - 1}$.  Discretisation of the distribution is represented by the gray rectangular areas in Fig.~\ref{fig:systematic_errors}.

\subsection{Normalisation error}
This error occurs because we need to load the truncated and discretised distribution into a quantum state.
The mass of a probability distribution function (on its domain) or the sum of all the points of a probability mass function are equal to 1. However, this is altered after truncation and discretisation.
From Eq.~\eqref{Eq:discretized_expectation}, we can thus define 
\begin{equation}
    \tilde{f}_{X}(x_{i}) = \frac{f_{X}(x_{i})\Delta}{\sum_{i=0}^{N-1}f_{X}(x_{i})\Delta},
\end{equation}
which is illustrated by the yellow rectangular areas in Fig.~\ref{fig:systematic_errors}, and $\sum_{i=0}^{N-1}f_{X}(x_{i})\Delta$ is not necessarily equal to 1. 
Since the probability distribution loaded by a quantum state is always normalised (i.e. the squared amplitudes for a quantum state sum to 1), we have to use the normalised distribution $\tilde{f}_{X}(x_{i})$.
Then, the normalisation error is defined as
\begin{eqnarray}
    \epsilon_{{\rm NoE}, n}(g(X)) &=& \abs{
        \sum_{i=0}^{N-1}g(x_{i})f_{X}(x_{i})\Delta - \sum_{i=0}^{N-1}g(x_{i})\tilde{f}_{X}(x_{i})
    }, \\ &=& \abs{
        \sum_{i=0}^{N-1}g(x_{i})\left(f_{X}(x_{i})\Delta - \tilde{f}_{X}(x_{i})\right)
    }.
\end{eqnarray}
This means that our next intermediate quantity is
\begin{equation}
    \mathbb{E}\left[g(X)\right]_{\rm No} := \sum_{i=0}^{N-1}g(x_{i})\tilde{f}_{X}(x_{i}).
\end{equation}
where the subscript ``No'' refers to ``normalisation''.

\subsection{State-preparation error}

The state preparation error introduced in this section is distinct from the aforementioned errors (which are all unavoidable when working on a digital computer). 
This error arises from the difference between a quantum state that could, in principle, be prepared, and the quantum state that is actually being prepared. 
Unlike in classical computing, where the values of probability mass must take some discrete value, amplitude encoding for a quantum state allows in principle a continuous range of values constrained to the overall normalisation.

We want to load the normalised probabilities $\tilde{f}_{X}(x_{i})$ into a quantum state as $\sum_{i=0}^{N-1}\sqrt{\tilde{f}_{X}(x_{i})}|x_i\rangle$, where $|x_i\rangle$ are the computational basis states and $N = 2^{n}$ for $n$ qubits.\footnote{Note that \emph{any} normalised state can in principle be prepared - but this may require an arbitrarily deep circuit.}
However, depending on the state-preparation method, this state cannot be loaded exactly; rather it will be loaded approximately.
Denoting the prepared state as
\begin{equation}\label{Eq:prepared_quantum_state}
    |\psi \rangle = \sum_{i=0}^{N-1}\sqrt{p(x_{i})}|x_i\rangle,
\end{equation}
we then define the state-preparation error as
\begin{eqnarray}
    \epsilon_{SE}(g(X)) &=& \abs{
        \sum_{i=0}^{N-1}g(x_{i})\tilde{f}_{X}(x_{i}) - \sum_{i=0}^{N-1}g(x_{i})p(x_{i})
    }, \\ &=& \abs{
        \sum_{i=0}^{N-1}g(x_{i})\left(\tilde{f}_{X}(x_{i}) - p(x_{i})\right)
    }.
\end{eqnarray}
Finally, we define the quantity that we \emph{actually} estimate in QMCI as
\begin{equation}
    \mathbb{E}\left[g(X)\right]_{S} := \sum_{i=0}^{N-1}g(x_{i})p(x_{i}),
\end{equation} 
where $S$ stands for ``state'', meaning that the expectation of the function applied is computed for the discrete distribution loaded by the quantum state.
Evaluating all errors along the way 
\begin{equation}
    \mathbb{E}\left[g(X)\right]  \to \mathbb{E}\left[g(X)\right]_{T}  \to \mathbb{E}\left[g(X)\right]_{D, n} \to \mathbb{E}\left[g(X)\right]_{No} \to \mathbb{E}\left[g(X)\right]_{S},
\end{equation} leads to an upper bound on the total systematic error, as the sum of all aforementioned errors:
\begin{equation}
    \abs{\mathbb{E}\left[g(X)\right]-\mathbb{E}\left[g(X)\right]_{S}} \leq \epsilon_{SE}(g(X)) + \epsilon_{NE, n}(g(X)) + \epsilon_{DE, n}(g(X)) + \epsilon_{TE}(g(X)).
\end{equation} Adding the errors in this manner provides a loose upper bound on the discrepancy between $\mathbb{E}\left[g(X)\right]$ and $\mathbb{E}\left[g(X)\right]_{S}$, and understanding how each of these influences the final computation is of practical relevance for applications.

\subsection{Operational errors}

Until now, all errors studied apply for any arbitrary function applied $g(\cdot)$. 
We introduce here the notion of operational errors associated with how the imperfect probability distribution propagates when functions are applied thereto. 
This is not something that can be done completely independently of the actual functions that are being applied, and thus this needs to be part of the analysis for a particular calculation.

The QMCI engine provides operations to be applied to the random variables represented in a quantum circuit.
We call such operations \emph{circuit enhancement operations}.
When we apply certain enhancement operations, these operations might introduce additional errors. 
In the QMCI engine, there are many enhancement operations (most of which are binary), and we here consider the thresholding operation, which we use for $\textrm{VaR}$ and $\cValueAtRisk$ calculations in this paper.
Let us say we want to estimate $\mathbb{E}\left[X\,\Theta(X \geq V_{Th})\right]$.
In the QMCI engine, we achieve this by defining a threshold with an inclusive upper bound of $V_{Th}$.
However, if $V_{Th} \in \left(x_{i}, x_{i + 1}\right)$ (open interval), then the engine uses $x_{i} $ as the bound of the threshold.
This gives rise to the thresholding error defined as
\begin{equation}
    \epsilon_{ThE}(X\,\Theta(X \geq V_{Th})) = \abs{
        \mathbb{E}\left[X\,\Theta(X \geq V_{Th})\right] - \mathbb{E}\left[X\,\Theta(X \geq x_{i})\right]
    }.
\end{equation} 
As mentioned previously, we use the thresholding operation for the $\ValueAtRisk$ and $\cValueAtRisk$ estimations in Section~\ref{sec:var-cvar-estimations}, and we explain the accuracy of the estimates based on the interplay between the discretisation and thresholding errors.
Therefore, in Section~\ref{sec:var-cvar-estimations}, we use an error that combines discretisation and thresholding errors into a single measure, denoted as $\epsilon_{DThE, n}$.
Having introduced all the systematic errors, in the next section, we apply our systematic error analysis to different use cases revealing insights into the performance of QMCI and its bottlenecks.

\section{Estimating (Conditional-)Value-at-Risk using the quantum Monte Carlo integration engine}
\label{sec:var-cvar-estimations}

In this section, we use the QMCI engine to estimate $\ValueAtRisk_{\alpha}$ and $\cValueAtRisk_{\alpha}$, and we benchmark the results against both the analytical values and classical estimations.
Using the QMCI engine's in-built state-preparation library we load a $5$-qubit Gaussian distribution, and then use this circuit to estimate $\ValueAtRisk_{\alpha}$ and $\cValueAtRisk_{\alpha}$. 
Performing QMCI using this circuit in general yields accurate estimates of the mean and second moment~\cite{akhalwaya2023modular}. 
However, we demonstrate that $\ValueAtRisk_{\alpha}$ can only be estimated accurately for small $\alpha$ values or for specific configurations (``sweet spots'') where systematic errors are minimised. 
Similarly, $\cValueAtRisk_{\alpha}$ estimates are only accurate at these specific ``sweet spots''.
Consequently, we conclude that encoding a Gaussian distribution using only $5$ qubits is insufficient for running SBO where $\alpha$ varies.
Based on our systematic error analysis, we demonstrate that the primary source of error in both $\ValueAtRisk_{\alpha}$ and $\cValueAtRisk_{\alpha}$ estimations is the thresholding error, which is related to the discretisation error.
Then, we determine the minimum number of qubits required to suppress these errors to a sufficient degree.
In other words, we show that, given a sufficient number of qubits (and assuming accurate state-preparation techniques when using a larger number of qubits), QMCI provides accurate estimations and exhibits the expected quadratic advantage in terms of RMSE scaling as compared to CMCI.

\subsection{Value-at-Risk estimations}\label{subsec:var_cvar_background}

Risk management constitutes a fundamental element within the framework of volatile economic environments. 
It encompasses the identification and analytical assessment of risk indicators, aimed at appraising and mitigating their detrimental effects.  This process is important in ensuring the stability and resilience of financial and economic systems against uncertainties inherent in market dynamics.

The significance of $\ValueAtRisk$ and $\cValueAtRisk$ as risk measures cannot be overstated, forming the foundation of the Basel III regulatory agreement \cite{basel3} (requiring credit granting institutions such as banks to perform stress tests using VaR-related measures). 
The computation of these risk measures, therefore, is of paramount importance, and practitioners utilise Monte Carlo simulations in order to determine the $\ValueAtRisk$ and $\cValueAtRisk$ of a given portfolio \cite{Hong2014}.  This process typically involves constructing a model that represents the portfolio's assets, and then calculating the aggregate value of the portfolio across $N$ distinct scenarios based on variations in the model's input parameters. 
$\ValueAtRisk_{\alpha}$ is attractive because it is a conceptually simple and intuitive quantification of risk.
More formally, we may represent the gains (loss if negative) of an investment portfolio over a fixed time as a random variable $X$. 
Given a probability level $0 < \alpha < 1$, $\ValueAtRisk_{\alpha}$ of the random variable $X$ is given by 
\begin{equation}
    \ValueAtRisk_{\alpha}(X) = \inf \{\gamma : \Pr(X\le \gamma) \ge \alpha \}. \label{eq:var}
\end{equation}
$\ValueAtRisk$ is equivalent to the  percentile for a probability distribution. 
We see that when the loss distribution is continuous, $\ValueAtRisk_{\alpha}(X)$ is the loss such that
\begin{equation}
    \Pr(X \leq \ValueAtRisk_{\alpha}(X)) = \E[\Theta(\ValueAtRisk_{\alpha}(X)-X)] = \alpha, \label{eq:var_est}
\end{equation}
where $\Theta(\cdot)$ is the indicator function as defined in \eqref{eq:indicator_var}. 
$\ValueAtRisk_{\alpha}$ is interpreted as the threshold value below which the probability of the portfolio or the investment losses exceeding the value is $\alpha$ or less.

\begin{algorithm}[h!]
\caption{The bisection algorithm for estimating $\ValueAtRisk_{\alpha}(X)$}
\label{alg:bisec}
 \nl Start with two values $a$ and $b$ such that $g(a)>\alpha$ and $g(b) < \alpha$\;
 \nl \While{$|b-a| \ge \epsilon$}
 {
 \nl Estimate $g(c)$, where $c = \frac{a+b}{2}$ is the the midpoint of the interval $(a,b)$\; \label{alg:comp_mid}

\nl \uIf{$g(c) = \alpha$} 
{
	Stop and return $\ValueAtRisk_{\alpha}(X) = c$\; 
}
\nl \uElseIf{$g(c) < \alpha $}
{
\nl  Set $b = c$\; 
}
\Else{
\nl Set $a = c$\;
}
}
\nl Return $\ValueAtRisk_{\alpha}(X) = \frac{a+b}{2}$\;		
\end{algorithm}

Despite the widespread use of $\ValueAtRisk_{\alpha}$, it lacks many important features that are crucial for risk management. 
Specifically, $\ValueAtRisk_{\alpha}$ does not look beyond the $\alpha$-percentile; that is, $\ValueAtRisk_{\alpha}$ does not account for losses exceeding it, and hence it can provide an inadequate picture of risks by failing to capture so-called ``tail risks''. 
Another challenge with $\ValueAtRisk_{\alpha}$ is that its optimisation becomes challenging when derived from scenario-based calculations (as is the case for the studies discussed in this work). 
Specifically, it has been shown \cite{771115,mckay1996var} that $\ValueAtRisk_{\alpha}$, as a function of portfolio allocations, can  possess several local extrema; this significantly increases the difficulty in identifying the most advantageous portfolio composition, or for accurately computing the $\ValueAtRisk_{\alpha}$ for a given portfolio mix.

Instead, $\cValueAtRisk$ has become the modern standard risk measure for a variety of practical applications, since $\cValueAtRisk_{\alpha}$ is recognized as a suitable model of risk in volatile economic circumstances. It is widely used for modeling and optimising hedge and credit risks \cite{kisiala2015conditional}. 
Furthermore, in terms of regulation, both in the insurance domain with the Solvency II Directive \cite{solvencyii} and in the financial domain with Basel III Rules, the role, importance, and applicability of $\cValueAtRisk_{\alpha}$ cannot be overstated.

While $\ValueAtRisk_{\alpha}$ gives a threshold level of loss with a specified confidence level, $\cValueAtRisk_{\alpha}$ goes a step further by quantifying the expected loss exceeding $\ValueAtRisk_{\alpha}$.
Practitioners are interested in $\cValueAtRisk_{\alpha}$ because it is easily expressible as an optimisation problem \cite{rockafellar2002conditional,rockafellar2000optimization,debrouwer2012maslowian}.
It is also continuous in $\alpha$ and convex (concave if seen from the insurance perspective) in the variables. 
The definition of $\cValueAtRisk_{\alpha}$, as an expectation, is given by
\begin{eqnarray}
\begin{aligned}
 \cValueAtRisk_{\alpha} (X) &= \E[X|X \ge \ValueAtRisk_{\alpha}(X)], \\ 
& = \frac{1}{1-\alpha} \int_{-\infty}^{+\infty} \Theta(x -\ValueAtRisk_{\alpha}(X)) x f(x) \rmd x.
 \label{eq:cvar}
 \end{aligned}
\end{eqnarray}

This definition can be extended to the discrete case \cite{rockafellar2002conditional}, where $\cValueAtRisk_{\alpha}$ is the weighted average of $\ValueAtRisk_{\alpha}$ and the expected losses strictly exceeding $\ValueAtRisk_{\alpha}$. 

While the preferred risk measure is $\cValueAtRisk_{\alpha}$, a reader might assume that $\ValueAtRisk_{\alpha}$ is not relevant. 
However, this is not the case since an interesting example of portfolio optimisation under uncertainty is one which calculates $\ValueAtRisk_{\alpha}$ and optimizes $\cValueAtRisk_{\alpha}$ simultaneously \cite{rockafellar2000optimization}. 
Interestingly, this SBO is one of the better candidates for a quantum solution, where one might hope to obtain maximal benefits by exploiting a quadratic speedup using QMCI.

\subsection{Estimation of $\ValueAtRisk_{\alpha}$ and $\cValueAtRisk_{\alpha}$ using Monte Carlo integration}\label{subsec:classical_var_cvar_estimators}
We now describe how to estimate $\ValueAtRisk_{\alpha}$ and $\cValueAtRisk_{\alpha}$ using CMCI. 
Estimation of $\ValueAtRisk_{\alpha}(X)$ can be expressed as a root-finding problem i.e.,  $g(\gamma) = \E[\Theta(\gamma-X)]  $, with $\gamma = \ValueAtRisk_{\alpha}(X)$.
We can use a simple bisection method to find the root of $g(\gamma) = \alpha$, as outlined in Algorithm~\ref{alg:bisec}.
The expectation value in the definition of the function $g(\gamma)$ is estimated using MCI (CMCI or QMCI).
Suppose that the random variable $X$ follows a distribution $f_{X}(x)$. 
Classically, we sample $\mathcal{S}$ random numbers from $f_{X}(x)$, denoted by the sorted observations $\{\tilde{X}_1,\cdots,\tilde{X}_{\mathcal{S}}\}$ such that  $\tilde{X}_1\le \cdots \le\tilde{X}_{\mathcal{S}}$.
Then, an estimate of $g(\gamma)$ can be obtained 
\begin{equation}\widehat{g(\gamma)} := \frac{I_{\alpha}}{\mathcal{S}},
\end{equation}
where $I_{\alpha} = \lfloor \alpha \mathcal{S} \rfloor$ denotes the index of the $I_{\alpha}$-th observation in $\tilde{X}$. 
Next, we turn to the estimation of $\cValueAtRisk_{\alpha}(X)$.
Following from \eqref{eq:cvar}, we can estimate $\cValueAtRisk_{\alpha}(X)$ as
\begin{equation}
    \widehat{\cValueAtRisk}_{\alpha}(X) := \frac{\tilde{X}_{ I_{\alpha}+1  }+\cdots+\tilde{X}_{\mathcal{S}}}{\mathcal{S}-I_{\alpha}}. \label{eq:discreteCVaR}
\end{equation}

\subsection{Numerical Experiments}\label{subsec:var-cvar-with-Gaussian}

\begin{table}[!htb]
    \centering
    \begin{tabular}{ |p{3cm}  || p{3cm}|   }
    \hline 
        \textbf{parameter} & \textbf{value / choice}  \\ \hline \hline 
        $X$ & $\mathcal{N}(\mu,\sigma^2)$\\ \hline
        $\mu$ & $0.10$\\ \hline
         $\sigma$ & $0.05$  \\  \hline
    \end{tabular}
    \caption{The parameters of the Gaussian distribution used for the $\ValueAtRisk_{\alpha}$ and $\cValueAtRisk_{\alpha}$ estimations.}
    \label{tab:gaussian-distribution-parameters}
\end{table}

We use a Gaussian distribution to model the underlying random variable $X$ (due to its known analytical solutions) to benchmark the performance of QMCI and CMCI, not only against one another, but also with respect to the closed-form solutions to the problems.
This approach allows us to make precise comparisons and assess the reliability of using QMCI in practical scenarios, ensuring that any observed discrepancies or improvements in the results obtained using QMCI can be accurately attributed to the method itself, rather than to characteristics of the distribution. 
These numerical analyses reveal that QMCI can provide similar estimates in terms of RMSE as CMCI for certain $\alpha$ values, however the numerical results also demonstrate that for particular $\alpha$ values QMCI gives large errors that lead to inaccurate estimates. 
In addition, based on $\cValueAtRisk_{\alpha}$ estimation, we also showcase the expected quadratic advantage of QMCI over CMCI for those particular configurations where errors are sufficiently suppressed. 
This section presents the results and describes how they were obtained, leaving the analysis and explanation of the observed results for Section~\ref{subsec:errors_var_cvar}.

Before presenting the results, we describe the distribution and its parameters, and provide analytical expressions for the quantities estimated using CMCI and QMCI.
For the analyses in this section, we use a one-dimensional Gaussian random variable $X \sim  \mathcal{N}(\mu,\sigma^2)$ with the parameters specified in Table~\ref{tab:gaussian-distribution-parameters}.
For the bisection method used to determine $\ValueAtRisk_{\alpha}$ (as described in Algorithm~\ref{alg:bisec}),  $g(\gamma)$ can be evaluated analytically as follows:
\begin{equation}
\label{eq:analytical-var-gaussian}
      g(\gamma) = \E[\Theta(\gamma-X) ]=\int_{-\infty}^\gamma \rmd x \, \frac{e^{-\frac{(x-\mu )^2}{2 \sigma ^2}}}{\sqrt{2 \pi } \sigma } =\frac{1}{2}\left(1+\erf(\frac{\gamma-\mu}{\sqrt{2}\sigma})\right). 
\end{equation}
 Given the value of $\ValueAtRisk_{\alpha}$, the $\cValueAtRisk_{\alpha}(X)$ in \eqref{eq:cvar} can similarly be analytically calculated for a Gaussian distribution. We omit this calculation for brevity.

\subsubsection{Estimation of $\ValueAtRisk_{\alpha}$ using the bisection method}
\label{subsubsec:var_estimate}

\begin{figure}[!t]
    \centering
    \includegraphics{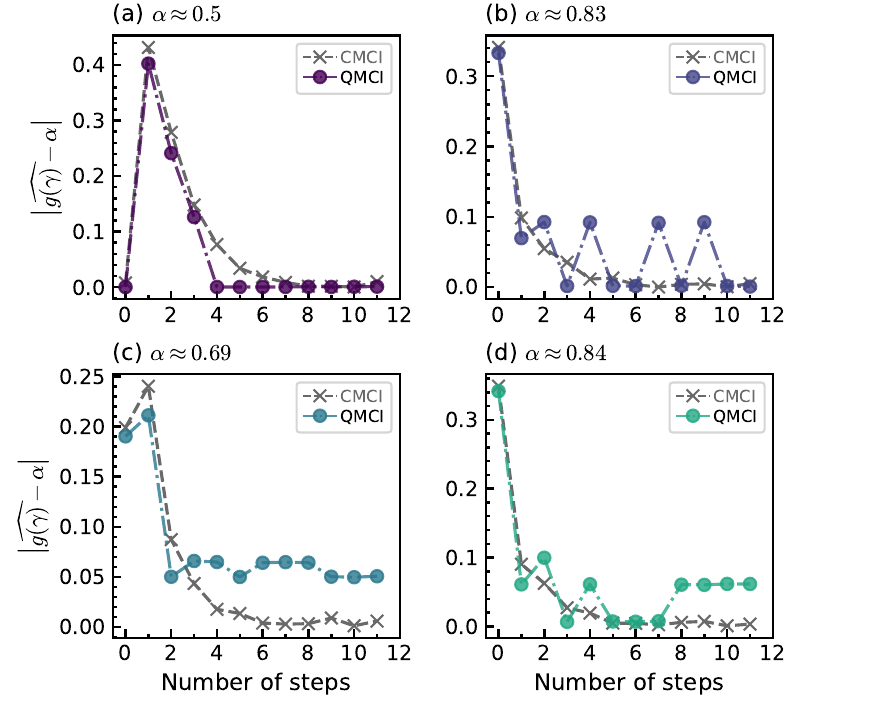}
    \caption{
    Difference between the analytical value of $\alpha$ and the estimate $\widehat{g(\gamma)}$  obtained from QMCI or CMCI at each iteration of the loop of {Algorithm~\ref{alg:bisec}} presented for four $\alpha$ values.
    In each iteration of the loop, $\widehat{g(\gamma)}$ are obtained using 10,000 samples  for both QMCI and CMCI.
    For (a)--(b), QMCI shows similar trends to CMCI, and both converge to similar values and have small errors at the last step,  where Algorithm~\ref{alg:bisec} reaches its stopping criterion.
    For (c)--(d), QMCI does not get close to the analytical value of $\alpha$ and has large errors at the last step.
    The $\ValueAtRisk_{\alpha}$ estimations at the last iteration and the corresponding analytical values are given in Table~\ref{tab:var_estimation_table}.
    }
    \label{fig:var_cost_value}
\end{figure}

\begin{table}[!ht]
    \centering
    \begin{tabular}{ |c||c|c|c| } 
    \hline \hline
         & Analytical value & CMCI & QMCI \\ \hline
         $\ValueAtRisk_{0.50}$ & 0.100 & 0.101 & 0.112 \\ \hline 
         $\ValueAtRisk_{\sim0.69}$ & 0.125 & 0.126 & \textcolor{red}{0.132}  \\  \hline 
         $\ValueAtRisk_{\sim0.83}$ & 0.148 & 0.149 & 0.148  \\  \hline 
         $\ValueAtRisk_{\sim0.84}$ & 0.150 & 0.150 & \textcolor{red}{0.165}  \\  \hline 
    \end{tabular}
    \caption{$\ValueAtRisk_{\alpha}$ estimations at the last iteration of the bisection method shown in Fig.~\ref{fig:var_cost_value} for QMCI and CMCI with 10,000 samples from the Gaussian distribution (with the parameters in Table~\ref{tab:gaussian-distribution-parameters}).
    For two $\alpha$ values, the $\ValueAtRisk_{\alpha}$ estimates obtained using QMCI (marked with red) are far away from their analytical values.
    For the other two $\alpha$ values, QMCI and CMCI give similar $\ValueAtRisk_{\alpha}$ estimates.
    These behaviours are explained by the error analysis described in Section~\ref{subsec:errors_var_cvar}.}
    \label{tab:var_estimation_table}
\end{table}

We run Algorithm~\ref{alg:bisec} with the stopping criterion $\epsilon = 1 \times 10^{-4}$ and use QMCI or CMCI as a sub-routine to estimate $g(c)$. Note that the stopping criterion is independent of the MCI sub-routine, and we compare the estimates at the last step where Algorithm~\ref{alg:bisec} hits the stopping criterion. In our analysis, we set the initial values of $a$ and $b$ to $a = \mu+3\sigma$ and $b = \mu-3\sigma$, satisfying $g(a)>0$ and $g(b)<0$.
As in the the rest of the paper, we use an implementation of the MLE-QAE algorithm for QMCI and load the Gaussian distribution using a $5$-qubit circuit (i.e., creating $32$ support points for the discretisation).  For the QMCI estimation of $\E[\Theta(c-X) ]$ in Algorithm~\ref{alg:bisec}, we use the thresholding functionality of the QMCI engine.

Figure~\ref{fig:var_cost_value} presents a comparative analysis of the estimation errors for both QMCI and CMCI using $10000$ samples (in each step of each run with different $\alpha$).
The absolute difference between the analytical value ($\alpha$) and the estimated value ($\widehat{g(\gamma)}$) is plotted over multiple steps of the loop of {Algorithm~\ref{alg:bisec}} for four different $\alpha$ values: $0.5$, $0.83$, $0.69$, and $0.84$.
Each subplot illustrates how the estimation error evolves as the number of steps increases, offering insights into the performance of both integration methods.

In Fig.~\ref{fig:var_cost_value}~(a)--(b), where $\alpha \approx 0.5$ and $\alpha \approx 0.83$, respectively, QMCI shows similar trends to CMCI, and both converge to similar values and have small errors at the last step.
However, in Fig.~\ref{fig:var_cost_value}~(c)--(d), with $\alpha \approx 0.69$ and $\alpha \approx 0.84$, respectively, QMCI does not get close to the analytical value of $\alpha$ and has large errors at the last step.
The $\ValueAtRisk_{\alpha}$ estimations at the last iteration and the corresponding analytical values are given in Table~\ref{tab:var_estimation_table}.
For two $\alpha$ values (corresponding to (c)--(d) in Fig.~\ref{fig:var_cost_value}), the $\ValueAtRisk_{\alpha}$ estimates obtained using QMCI (marked with red) are not close to the analytical values.
For the other two $\alpha$ values, QMCI and CMCI give similar $\ValueAtRisk_{\alpha}$ estimates.

Overall, the results indicate that while QMCI has the potential to provide accurate estimations for these quantities, this is highly dependent on the specific conditions -- specifically the value of $\alpha$. These behaviours are explained by the error analysis in Section~\ref{subsec:errors_var_cvar}.

\subsubsection{Scaling of the root mean-squared error for $\cValueAtRisk_{\alpha}$ estimation}

In Section~\ref{subsubsec:var_estimate} we showed that (albeit only for certain $\alpha$ values) QMCI can produce accurate $\ValueAtRisk_{\alpha}$ estimates. However we did not analyse the convergence properties. 
In this section, based on $\cValueAtRisk_{\alpha}$ estimations, we investigate whether QMCI exhibits a quadratic advantage in RMSE scaling over CMCI, and we estimate $\cValueAtRisk_{\alpha}$ using QMCI and CMCI across different sample sizes and estimate their RMSE from 250 independent runs of $\cValueAtRisk_{\alpha}$ estimations.
We show that QMCI has a clear quadratic advantage for the same $\alpha$ values for which QMCI gave accurate $\ValueAtRisk_{\alpha}$ estimations. We explain these observations based on the error analysis in Section~\ref{subsec:errors_var_cvar}.

\begin{figure}[!t]
    \centering
    \includegraphics{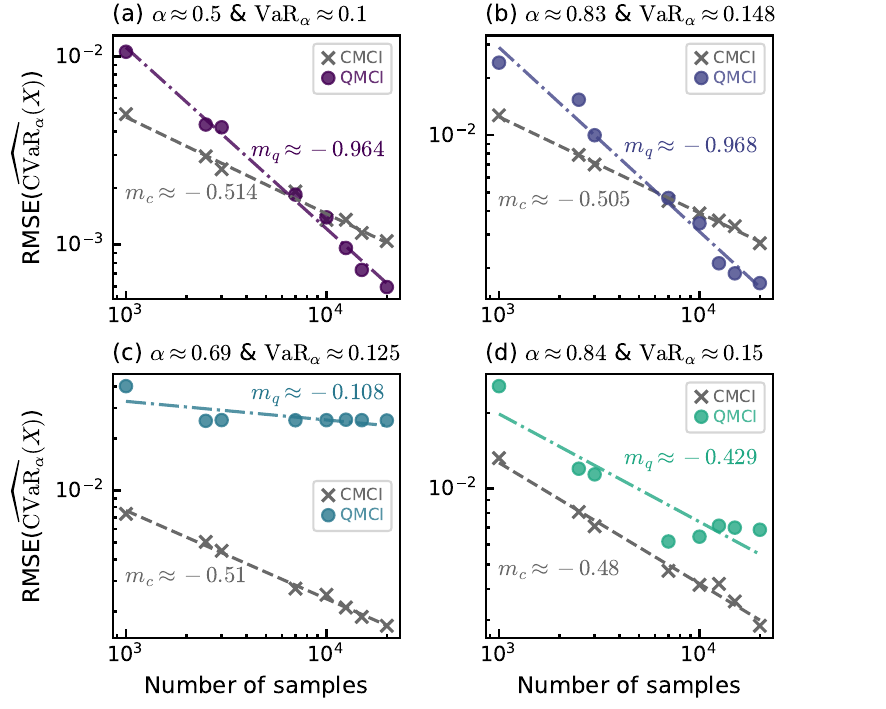}
    \caption{
    Comparisons of the RMSE scaling (as a function of number of samples $\mathcal{S}$) in $\cValueAtRisk_{\alpha}$ estimations using QMCI and CMCI for four $\alpha$ values.
    RMSE estimations are obtained by averaging over 250 independent runs.
    For (a)--(b), QMCI clearly performs better than CMCI, meaning $m_{q} \sim -1$ and $m_{c} \sim -0.5$, thus QMCI gives a better RMSE than CMCI for a sufficiently large number of samples.
    For (c), QMCI scales much worse than CMCI, while QMCI scaling is similar to CMCI for (d).
    }
    \label{fig:cvar_rmse}
\end{figure}

\begin{table}[!t]
    \centering
    \begin{tabular}{ |c||c|c| } 
    \hline \hline
         & RMSE for CMCI & RMSE for QMCI \\ \hline
         $\cValueAtRisk_{0.50}$ & 0.0010 & 0.0006 \\ \hline 
         $\cValueAtRisk_{\sim0.69}$ & 0.0016 & \textcolor{red}{0.0254}  \\  \hline  
         $\cValueAtRisk_{\sim0.83}$ & 0.0027 & 0.0017  \\  \hline 
         $\cValueAtRisk_{\sim0.84}$ & 0.0028 & \textcolor{blue}{0.0069}  \\ 
     \hline
    \end{tabular}
    \caption{RMSE of the $\cValueAtRisk_{\alpha}$ estimations using QMCI and CMCI with 20000 samples (for 250 independent runs with the Gaussian distribution in Table~\ref{tab:gaussian-distribution-parameters}).
    For $\alpha\sim 0.69$, the RMSE of QMCI (marked with red) is about an order of magnitude larger than CMCI.
    For $\alpha\sim 0.84$, the RMSE of QMCI (marked with blue) is larger than CMCI (but approximately the same order of magnitude).
    For the other two $\alpha$ values, QMCI has a smaller RMSE than CMCI.
    }
    \label{tab:cvar_rmse_table}
\end{table}

Figure~\ref{fig:cvar_rmse} presents comparisons of the RMSE scaling (as a function of number of samples) in $\cValueAtRisk_{\alpha}$ estimations using QMCI and CMCI for four $\alpha$ values.
Representing the number of samples with $\mathcal{S}$, we hereby denote the RMSE scaling as $\mathcal{O}(\mathcal{S}^{m_{q}})$ and $\mathcal{O}(\mathcal{S}^{m_{c}})$ for QMCI and CMCI, respectively.
In Fig.~\ref{fig:cvar_rmse}, we fit the RMSE estimations to obtain estimates of $m_{q}$ and $m_{c}$.
In Fig.~\ref{fig:cvar_rmse}~(a)--(b), QMCI clearly outperforms CMCI, meaning the the RMSE scales as $m_{q} \sim -1$ for QMCI, while $m_{c} \sim -0.5$ for CMCI.
Thus, in Fig.~\ref{fig:cvar_rmse}~(a)--(b), QMCI has better RMSE scaling than CMCI for a sufficiently large number of samples.
In Fig.~\ref{fig:cvar_rmse}~(c), the RMSE for QMCI is almost constant as a function of the number of samples, with much worse scaling than for CMCI (which has $m_{c} \sim -0.5$ for all cases).
Finally, the QMCI RMSE scaling is similar to that of CMCI in Fig.~\ref{fig:cvar_rmse}~(d).
The RMSE values for $\mathcal{S}=20000$ are given in Table~\ref{tab:cvar_rmse_table}.

In summary, Fig.~\ref{fig:cvar_rmse} reveals that QMCI achieves a clear quadratic speedup over CMCI for those $\alpha$ values where good $\ValueAtRisk_{\alpha}$ estimates are obtained in the previous section. 
Conversely, for the other two $\alpha$ values whereby QMCI does not provide accurate $\ValueAtRisk_{\alpha}$ estimates, the RMSE scaling fails to show any advantage, and is worse than for CMCI.
The underlying causes of these behaviors are explained in Section~\ref{subsec:errors_var_cvar}.

\subsection{Systematic error analyses for $\ValueAtRisk_{\alpha}$ and $\cValueAtRisk_{\alpha}$ estimation}\label{subsec:errors_var_cvar}

In Section~\ref{subsec:var-cvar-with-Gaussian}, we performed CMCI and QMCI computations to estimate the $\ValueAtRisk$ and $\cValueAtRisk$ for four different $\alpha$ values, and demonstrated that QMCI provides both accurate estimates and an advantage in RMSE scaling for only two of these values. 
In this section, we analyse the systematic errors present in QMCI in order to understand the reasons behind this. 
Specifically, we introduce the concept of ``scaled error'', which is the ratio of a particular systematic error to the analytical value.
This analysis is critical in order to explain the observed variations in QMCI performance.

\begin{figure}[!t]
    \centering
    \includegraphics[width=0.9\linewidth]{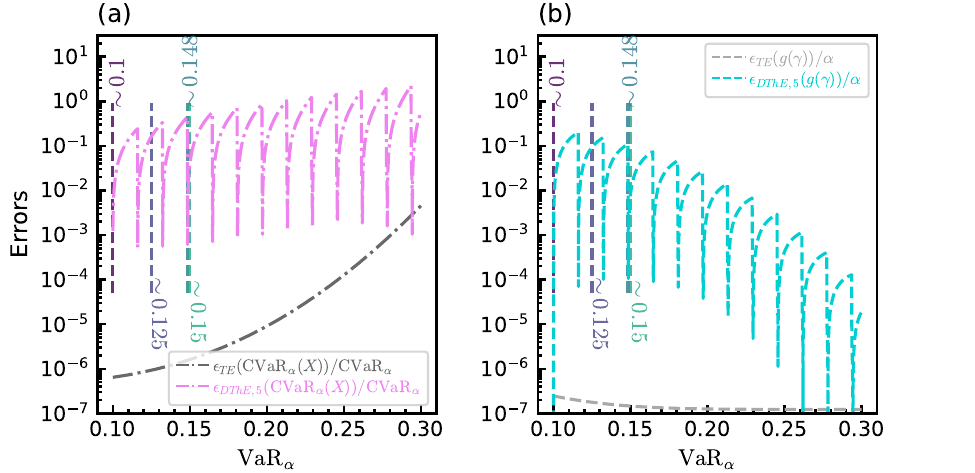}
    \caption{
    The ratio of $\epsilon_{DThE, 5}$ (the combined discretisation and thresholding error) and the truncation errors to the analytical values as a function of $\ValueAtRisk_{\alpha}$ for (a) $\cValueAtRisk_{\alpha}$ and (b) $g(\gamma)$ estimations with the 5-qubit Gaussian distribution (with the parameters given in Table~\ref{tab:gaussian-distribution-parameters}).
    $\cValueAtRisk_{\alpha}(X)$ and $g(\gamma)$ represent the function applied, and $\cValueAtRisk$ and $\alpha$ (i.e., without the random variable $X$) represent the analytical values.
    The four $\ValueAtRisk_{\alpha}$ values marked with the vertical dashed lines correspond to the $\alpha$ values considered in Tables~\ref{tab:var_estimation_table}--\ref{tab:percent_error_table}.
    The exact values of the scaled errors for the marked $\ValueAtRisk_{\alpha}$ values are given in Table~\ref{tab:percent_error_table}.
    }
    \label{fig:percent_errors}
\end{figure}

We begin by defining the systematic errors for $\cValueAtRisk_{\alpha}$ estimation, noting that similar definitions for $\widehat{g(\gamma)}$ are omitted for brevity. 
We then present the behavior of these errors in Fig.~\ref{fig:percent_errors} as a function of $\ValueAtRisk_{\alpha}$ and the $\ValueAtRisk_{\alpha}$ values corresponding to the four $\alpha$ values analysed previously are marked with vertical lines.
Table~\ref{tab:percent_error_table} lists the scaled errors for the four $\alpha$ cases, providing a quantitative comparison. 
Finally, we examine the maximum of the scaled errors as a function of the number of qubits, to determine the qubit count required to suppress these errors below an empirically determined upper bound.

\begin{table}[!htb]
    \centering
    \begin{tabular}{ |c||c|c| } 
    \hline \hline
         & $\epsilon_{DThE, 5}(g(\gamma))/\alpha$ & $\epsilon_{DThE, 5}(\cValueAtRisk_{\alpha}(X))/\cValueAtRisk_{\alpha}$ \\ \hline
         $\ValueAtRisk_{0.50} = 0.1$ & $2.73\times 10^{-8}$ & 0.001 \\ \hline 
         $\ValueAtRisk_{\sim0.69} = 0.125$ & \textcolor{red}{0.093} & \textcolor{red}{0.162}  \\  \hline  
         $\ValueAtRisk_{\sim0.83} \sim 0.148$ & 0.001 & 0.003  \\  \hline 
         $\ValueAtRisk_{\sim0.84} = 0.15$ & \textcolor{red}{0.008} & \textcolor{blue}{0.039}  \\  \hline 
    \end{tabular}
    \caption{The ratios of $\epsilon_{DThE, 5}$ (the combined discretisation and thresholding error) to their analytical values for $\cValueAtRisk_{\alpha}$ and $g(\gamma)$ estimations with the $5$-qubit Gaussian distribution (with the parameters given in Table~\ref{tab:gaussian-distribution-parameters}).
    For $\alpha=0.5$ and $\alpha\sim 0.83$, the error ratios for the $\cValueAtRisk_{\alpha}$ and $g(\gamma)$ estimations are of the order of (or smaller than) $10^{-3}$, and these are the same $\alpha$ values that give accurate results in Tables~\ref{tab:var_estimation_table}--\ref{tab:cvar_rmse_table} and show better RMSE scaling in Fig.~\ref{fig:cvar_rmse}.
    For $\alpha\sim 0.69$ and $\alpha\sim 0.84$, the error ratios for the $g(\gamma)$ estimations (marked with red) are, respectively about two orders and an order of magnitude larger than $10^{-3}$, and these are the $\alpha$ values for which QMCI give worse estimates than CMCI in Table~\ref{tab:var_estimation_table}.
    For $\alpha\sim 0.83$, the error ratio for $\cValueAtRisk_{\alpha}$ is an order of magnitude larger than $10^{-3}$, and the RMSE values of CMCI and QMCI for this $\alpha$ value in Table~\ref{tab:cvar_rmse_table} are comparable to each other.
    Finally, the error ratio for $\cValueAtRisk_{\alpha}$ for $\alpha\sim 0.69$ is two orders of magnitude larger than $10^{-3}$, and the RMSE values of QMCI for this $\alpha$ value in Table~\ref{tab:cvar_rmse_table} are an order of magnitude larger than the RMSE values of CMCI.\\}
    \label{tab:percent_error_table}
\end{table}

Our findings show that for the two $\alpha$ values yielding good estimates, the scaled errors are of the order of $10^{-3}$ or smaller. 
For the other two $\alpha$ values, the scaled errors exceed $10^{-3}$.
Furthermore, we demonstrate that for a Gaussian distribution truncated to $\left[-6\sigma, 6\sigma\right]$ and loaded using a circuit with greater than or equal to 15 qubits, then the maximal error ratios are all less than $10^{-3}$ for $\cValueAtRisk_{\alpha}$ estimation and for $\ValueAtRisk_{\alpha}$ within the range $\left(-\infty, 4\sigma\right]$. 
This indicates that increasing the number of qubits can effectively reduce systematic errors, thereby enhancing the accuracy of QMCI estimates for a broader range of $\alpha$ values.
However, this assumes the existence of state-preparation methods that can accurately and efficiently load such distributions.

\subsubsection{Systematic errors for $\cValueAtRisk_{\alpha}$ estimation}
\begin{figure}[t]
    \centering
    \includegraphics{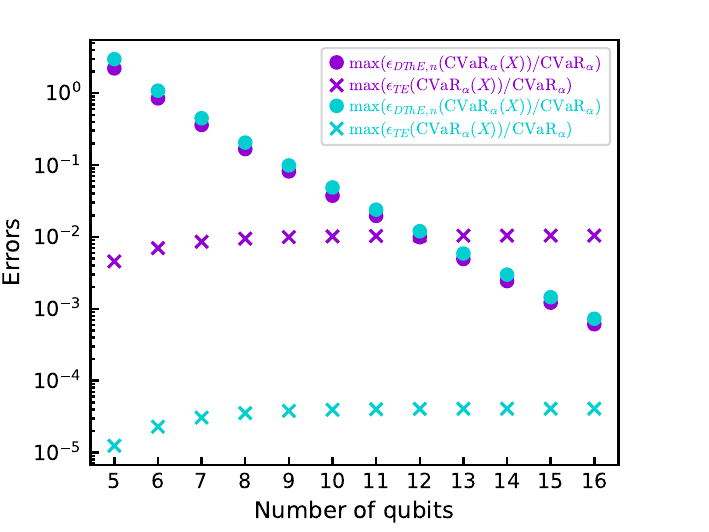}
    \caption{
    Maximum of the ratio of $\epsilon_{DThE, 5}$ (the combined discretisation and thresholding errors) to analytical values for $\cValueAtRisk_{\alpha}$ estimations for $\ValueAtRisk_{\alpha}$ in the range $\left(-\infty, 4\sigma\right]$ with the 5-qubit Gaussian distribution (with the parameters given in Table~\ref{tab:gaussian-distribution-parameters}).
    Different colours represent different truncation ranges.
    Purple (turquoise) represents the Gaussian distribution with range $\left[-5\sigma, 5\sigma\right]$ ($\left[-6\sigma, 6\sigma\right]$).
    }
    \label{fig:percent_errors_for_n_qubits}
\end{figure}

We now provide concrete expressions for the truncation error and the combined discretisation and thresholding errors in $\cValueAtRisk_{\alpha}$ estimation. This exemplifies how we consider concrete error expressions tailored for a specific function applied.
Additionally, Figs.~\ref{fig:percent_errors}~-~\ref{fig:percent_errors_for_n_qubits} show the ``scaled errors'', which we prefer for meaningful comparisons of the errors in different regimes. However, this is only possible when the analytical value or a ground truth for the estimated quantity is known.\\

For a given (inclusive) lower threshold value $\textrm{VaR}_{\alpha} = \gamma$, $\cValueAtRisk_{\alpha}$ is given by
\begin{equation}
  \cValueAtRisk_{\alpha}(X) = \mathbb{E}\left[X | X \geq \gamma \right] = \frac{1}{\Pr(X \geq \gamma)}\int_{\gamma}^{\infty}xf(x)dx \, ,
\end{equation}
and we write the truncated conditional expectation as
\begin{equation}
  \mathbb{E}\left[X | X \geq \gamma \right]_{T} = \frac{1}{\Pr(X \geq \gamma)}\int_{\gamma}^{x_{u} + \Delta/2}xf(x)dx \, ,
\end{equation}
where we assume that $\Pr(X \geq \gamma)$ is known (the same for all following definitions).
The truncation error then becomes
\begin{equation}
  \epsilon_{TE}(\cValueAtRisk_{\alpha}(X)) := \epsilon_{TE}(\mathbb{E}\left[X | X \geq \gamma \right]) = \lvert \mathbb{E}\left[X | X \geq \gamma \right] - \mathbb{E}\left[X | X \geq \gamma \right]_{T} \rvert \, .
\end{equation}
Next, we write the discretised conditional expectation as
\begin{equation}
  \mathbb{E}\left[X | X \geq \gamma \right]_{DTh, n} = \frac{1}{\Pr(X \geq \gamma)}\sum_{i=I_{\alpha}}^{N-1}x_{i}f(x_{i})\Delta \, ,
\end{equation}
where $I_{\alpha}$ is the index of the support point corresponding to $\gamma + \cfrac{\Delta}{2}$, and $N = 2^n$ is the number of points in the support for $n$ qubits.
If $\gamma + \cfrac{\Delta}{2}$ is exactly equal to a support point, $\mathbb{E}\left[X | X \geq \gamma \right]_{DTh, n} = \mathbb{E}\left[X | X \geq \gamma \right]_{D, n}$.
Otherwise, $\mathbb{E}\left[X | X \geq \gamma \right]_{DTh, n}$ also includes the thresholding error.
Finally, we represent the combined discretisation and thresholding errors as
\begin{equation}
  \epsilon_{DThE, n}(\mathbb{E}\left[X | X \geq \gamma \right]) = \lvert \mathbb{E}\left[X | X \geq \gamma \right]_{T} - \mathbb{E}\left[X | X \geq \gamma \right]_{DTh, n} \rvert \, .
\end{equation}
Similarly, starting from \eqref{eq:var_est}, we can write the intermediate quantities and errors for $g(\gamma)$.

\subsubsection{Numerical analysis of the systematic errors in $\ValueAtRisk_{\alpha}$ and $\cValueAtRisk_{\alpha}$ estimations}

Fig.~\ref{fig:percent_errors} shows the ratio of the combined discretisation and thresholding error ($\epsilon_{DThE, 5}$) and truncation errors to the analytical values as a function of $\ValueAtRisk_{\alpha}$ for (a) $\cValueAtRisk_{\alpha}$ and (b) $g(\gamma)$ estimations.  
The functions $\cValueAtRisk(X)$ and $g(\gamma)$ represent the estimated values, with their analytical values represented by $\cValueAtRisk$ and $\alpha$ (i.e., without the random variable $X$). 
Fig.~\ref{fig:percent_errors} shows that the truncation error ratio is almost always smaller than the combined discretisation and thresholding error.
We observe that $\epsilon_{DThE, 5}$ exhibits ``troughs'' at certain $\ValueAtRisk_{\alpha}$ values.
This is explained by the increase of the thresholding error when the threshold $\gamma$ is shifted from a point that is exactly between two support points to another mid-point.
$\epsilon_{DThE, 5}$ reaches its maximum just before $\gamma$ equals the next mid-point, and the size of the error is represented in Fig.~\ref{fig:systematic_errors}. 
For the four aforementioned values of $\alpha$, we report the values of the different ratios error to the analytical values in Table~\ref{tab:percent_error_table}.

Table~\ref{tab:percent_error_table} shows that, for $\alpha \approx 0.69$ and $\alpha \approx 0.84$, $\epsilon_{DThE, 5}(g(\gamma))/\alpha$ values are significantly larger than $10^{-3}$, correlating with the poorer QMCI performance compared to CMCI for the estimation of $g(\gamma)$ for these $\alpha$ values in Table~\ref{tab:var_estimation_table}. 
In contrast, for $\alpha = 0.5$ and $\alpha \approx 0.83$, $\epsilon_{DThE, 5}(g(\gamma))/\alpha$ values are of the order of or smaller than $10^{-3}$, and QMCI provided accurate estimations of $g(\gamma)$ for these $\alpha$ values in Table~\ref{tab:var_estimation_table}.
Similarly, RMSE scaling for $\cValueAtRisk_{\alpha}$ estimation in Fig.~\ref{fig:cvar_rmse} demonstrates a quadratic speed up for $\alpha = 0.5$ and $\alpha \approx 0.84$, and Table~\ref{tab:percent_error_table} shows that the $\epsilon_{DThE, 5}(\cValueAtRisk_{\alpha}(X))/\cValueAtRisk_{\alpha}$ values for these $\alpha$ values are again of the order of $10^{-3}$.
For $\alpha\sim 0.84$, $\epsilon_{DThE, 5}(\cValueAtRisk_{\alpha}(X))/\cValueAtRisk_{\alpha}$ value is an order of magnitude larger than $10^{-3}$, and the RMSE values of CMCI and QMCI for this $\alpha$ value in Table~\ref{tab:cvar_rmse_table} are of the same order, with a similar RMSE scaling in Fig.~\ref{fig:cvar_rmse}.
Finally, $\epsilon_{DThE, 5}(\cValueAtRisk_{\alpha}(X))/\cValueAtRisk_{\alpha}$ for $\alpha\sim 0.69$ is two orders of magnitude larger than $10^{-3}$, and the RMSE values of QMCI for this $\alpha$ value in Table~\ref{tab:cvar_rmse_table} is an order of magnitude larger than the RMSE value of CMCI.
To summarise, we observe that if the scaled errors are below $10^{-3}$, QMCI provides accurate estimates and exhibits quadratic advantage.

\subsubsection{Maximum of the scaled errors for $\cValueAtRisk_{\alpha}$ estimation}
Given the empirical upper bound (of $10^{-3}$) for the scaled error, we now analyse the scaled errors for $\cValueAtRisk_{\alpha}$ estimation to determine the number of qubits required to keep systematic errors \emph{sufficiently suppressed} in order that QMCI can provide accurate estimations.
In particular, we show in Fig.~\ref{fig:percent_errors_for_n_qubits} the maximum of the ratio of $\epsilon_{DThE, 5}$ to analytical values for $\cValueAtRisk_{\alpha}$ estimations for $\ValueAtRisk_{\alpha}$ in the range $\left(-\infty, 4\sigma\right]$ for the Gaussian distribution. 
We analyse two different truncation ranges $\left[-5\sigma, 5\sigma\right]$ and $\left[-6\sigma, 6\sigma\right]$.
Fig.~\ref{fig:percent_errors_for_n_qubits} shows that the maximum of ratio truncation error for $5\sigma$ truncation is never $\lesssim 10^{-3}$.
However, the $6\sigma$ truncation -- despite having a slightly larger $\epsilon_{DThE, 5}$ error ratio -- achieves \emph{sufficiently small} truncation error ratios.
In both cases the $\epsilon_{DThE, 5}$ error ratios reach \emph{sufficiently small} values (i.e. $\lesssim 10^{-3}$) when the number of qubits is greater than or equal to $15$.

Our error analyses show that QMCI can provide accurate $\ValueAtRisk_{\alpha}$ and $\cValueAtRisk_{\alpha}$ estimates when all the scaled errors are kept small which happens when a sufficient number of qubits are used to load the distribution. 
To our knowledge, the determination of this number in the context of specific computations has not been previously explored to any significant degree.

\section{Mean-CVaR simulation-based portfolio optimisation over stable distributions} \label{sec:mean-cvar-SBO}

In 1952,  Roy formulated a safety-first criterion \cite{roy1952safety}, where an investor minimises the probability of wealth falling below some minimal level. The problem formulated by Roy is equivalent to optimising for a quantile, and hence equivalent to replacing the Variance by the $\cValueAtRisk$ -- see for example \cite{debrouwer2012maslowian}. We choose in this Section to focus on  the problem of portfolio optimisation formulated by minimising risk (the $\cValueAtRisk_{\alpha}$ specifically) while requiring a minimum expected return -- see also Ref.~\cite{Rockafellar2000OptimizationOC}. An ``equivalent'' formulation amounts to maximising the return by restricting large risks. This can be done by swapping the $\cValueAtRisk_{\alpha}$ and the expected return in the problem formulation and minimising the negative expected return with a penalised $\cValueAtRisk_{\alpha}$ constraint.

 Consider  a  portfolio of $n$ assets with a random return that minimises negative returns (penalising risk) together with budget constraints and bounds on each exposure. Given returns of $n$ assets, where $w_i$ is the weight of the $i$-th asset in the portfolio, and $X_i$ represents the return of asset $i$, where $i=1,\ldots,n$, then $\{ X_1,\cdots,X_n \} \sim \Omega_{X}$ and $w = (w_1,\cdots,w_n)$ are the decision variables.
The expected return of the portfolio over all assets is then $\E[X=\sum_{i=1}^n w_i X_i]$. Following insights from Ref.~\cite[Theorems 3 and 5]{Krokhmal2001}, we model the risky portfolio optimisation problem as follows:
\begin{mini}
{w \in  \mathbb{R}^{n}}{f(w,X) = -\E[X] + \lambda \cValueAtRisk_{\alpha}(X)}
{}{}
\addConstraint{w_{\min} \leq w_i \leq w_{\max},}{ \quad \forall i=1,\ldots,n}
\addConstraint{ \sum_{i=1}^{n} {w_i} =1; }{ }
\label{eq:mean-cvar-opt}
\end{mini}
where $w_{\min}$ and $w_{\max}$ are the minimum and maximum allowable weight for each asset, respectively. The constraint $\sum_{i=1}^{n} {w_i} =1$ implies no leverage. Furthermore, we assume no short positions, that is $w_i >0$ for all $i=1,\ldots,n$. The problem formulated in \eqref{eq:mean-cvar-opt} can be solved using existing solvers such as the Sequential Least-Squares Programming (SLSQP) algorithm \cite{Boggs1995sequential,Nocedal2006quadratic} that is a well-known method for solving constrained nonlinear optimization in an iterative manner, in which the objective function $f(w,X)$ can be evaluated using CMCI or QMCI. More explicitly, we give the schematic pseudo-code to solve the problem in {Algorithm~\ref{alg:formal_portfolio_optimisation_cvar}}.

\begin{algorithm}[ht!]
\caption{Iterative $\cValueAtRisk_{\alpha}$ Portfolio optimisation}
\label{alg:formal_portfolio_optimisation_cvar}
 \nl \textbf{Input:} Joint distribution of asset returns $\{X_i,i=1,\cdots,n\} \sim \Omega_X$, risk aversion parameter $\lambda$, confidence level $\alpha$\;
 \nl \textbf{Output:} Optimized portfolio weights ${w}^* \in [0,1]^n$\;
 \nl Initialize ${w}^{(0)}$ randomly subject to $\sum_i w_i^{(0)} = 1$\;
 \nl Define objective function $f(w,X) = - \E[{w}_i^\intercal X_i] + \lambda  \cValueAtRisk_{\alpha}(w_i ^\intercal X_i)$\;
 \nl $k \gets 0$\;
 \nl \Repeat{convergence criterion is met}{
 \nl     Simulate portfolio returns $X^{(k)} = \sum_{i=1}^n w_i^{(k)} X_i$ using CMCI or QMCI\;
 \nl     Calculate $\cValueAtRisk_{\alpha}$ for $X^{(k)}$ as $\cValueAtRisk_{\alpha}(X^{(k)})$ using CMCI or QMCI\;
 \nl     Update objective function $f^{(k)}(w,X) = -\E[X^{(k)}]  + \lambda  \cValueAtRisk_{\alpha}(X^{(k)})$\;
 \nl     Solve ${w}^{(k+1)} = \arg \min_{{w}} f^{(k)}({w,X})$ subject to $\sum_i w_i = 1$ and $w_{\min} \leq w_i \leq w_{\max}$\; 
 \nl     $k \gets k + 1$\;
 }
 \nl \Return ${w}^* = {w}^{(k)}$\;		
\end{algorithm}

\begin{table}[!t]
    \centering
    \begin{tabular}{|p{4cm} || p{3.5cm} | }
    \hline  
        \textbf{parameter} & \textbf{value / choice}  \\\hline \hline 
        $X_i$ & $\mathcal{N}(\mu_i,\sigma_i^2)$\\\hline 
        $\mu_1$ & $0.10$\\\hline 
        $\mu_2$ & $0.10$\\\hline 
         $\sigma_1$ & $0.05$  \\\hline 
         $\sigma_2$ & $0.10$  \\\hline 
         $\alpha$ & $0.95$\\\hline 
         $w_{\min}$ & $0.10$\\\hline 
         $w_{\max}$ & $0.90$\\\hline 
         $\lambda$ & $0.10$\\\hline
         number of samples & 10000\\\hline
    \end{tabular}
    \caption{Parameter values for the Mean-$\cValueAtRisk$ optimisation.}  
    \label{table:3rd-numerical-exp-params}
\end{table}

\subsection{Numerical Experiments}
\label{subsec:qexp_gaussian-levy-mean-cvar}

 The Mean-$\cValueAtRisk$ optimisation formulated in \eqref{eq:mean-cvar-opt} involves vector optimisation over uncertainty and the calculation of the expected return and $\cValueAtRisk$ in terms of multiple assets depend on the weight factor, thus the optimal solution cannot be obtained analytically. The numerical experiments discussed in this section therefore focus on how using QMCI as a subroutine affects the behavior and efficiency of the optimisation algorithm when compared to using CMCI as a subroutine. The numerical experiments will provide a playground to observe the difference in the convergence to the final optimised objective value when using QMCI and CMCI.

We consider the problem where the returns of the $i$-th asset $X_i, i=1,\cdots,n$ follows some specified distribution. For instance,  $X_i \sim \mathcal{L}(\mu_i,c_i)$ or $X_i \sim \mathcal{N}(\mu_i,\sigma_i^2)$, for each asset $i$, with $i=1,\cdots,n$. For simplicity, we choose a portfolio consisting of $n=2$ different assets, where the returns $X_1$ and $X_2$ are independent and Gaussian distributed with $X_i \sim \mathcal{N}(\mu_i,\sigma_i^2)$, i=1,2. The objective function \eqref{eq:mean-cvar-opt} is then expanded as follows
\begin{equation}
\label{eq:cvar_opt_gaussian}
   f(w,X) =  - \E[w_1 X_1 + w_2 X_2] + \lambda \cValueAtRisk_{\alpha}(w_1 X_1 + w_2 X_2).
\end{equation} 

We refer to Table~\ref{table:3rd-numerical-exp-params} for the values of all parameters chosen for the numerical experiments. The computation of $\cValueAtRisk_{\alpha}$ using both CMCI and QMCI follow the same procedure introduced in Section~\ref{sec:var-cvar-estimations}.  In the numerical experiments, all weights $w_i$ are initialised as $w_i^{(0)} = \frac{1}{n}$ for {Algorithm~\ref{alg:formal_portfolio_optimisation_cvar}}  used to solve the optimisation problem \eqref{eq:cvar_opt_gaussian}.

\begin{figure}[!ht]
    \centering
    \includegraphics{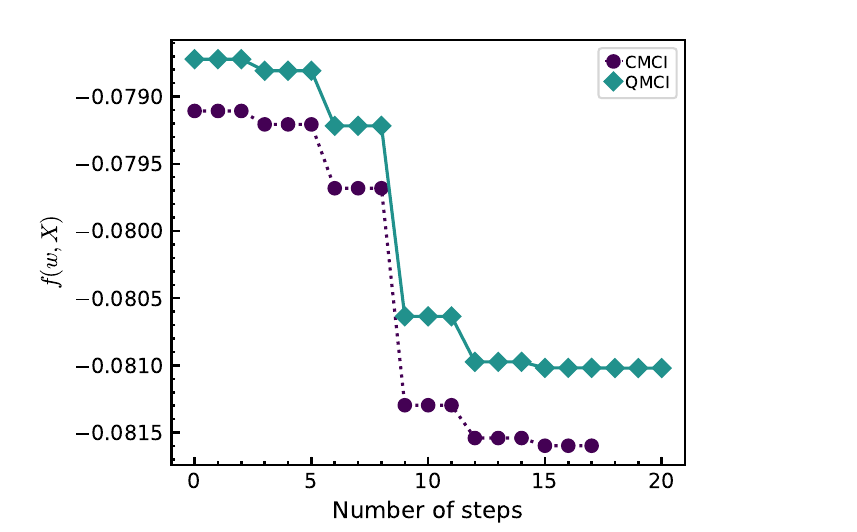}
    \caption{The estimated objective $f(w,X)$ versus the number of iterations in {Algorithm~\ref{alg:formal_portfolio_optimisation_cvar}} using both QMCI and CMCI. The final optimised objective for CMCI is $f(w,X) = -0.0807$ with optimised weights $w^* = [0.7894,0.2106]$, while for QMCI $f(w^*,X) = -0.0811$ with $w^* = [0.7245,0.2755]$. 
    }
    \label{fig:sub_fvssteps}
\end{figure}

\begin{figure}[!ht]
    \centering
    \includegraphics{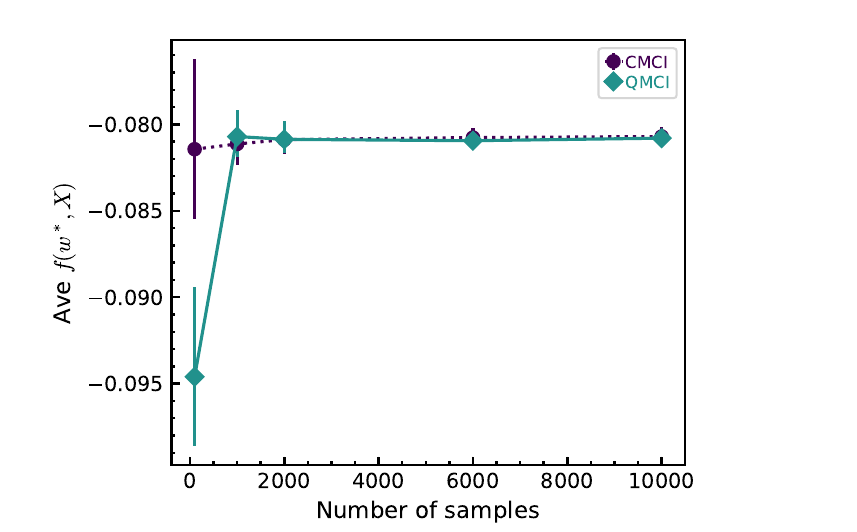}
    \caption{The average value of the optimised objective $f(w^*,X)$ as a function of the number of samples, in which each point of  $\bar{f}(w^*,X)$ is obtained by averaging over 100 independent runs.
    }
    \label{fig:sub_fvssamples}
\end{figure}

Figure~\ref{fig:sub_fvssteps} displays the values of the objective function $f(w,X)$ in the iterative procedure of {Algorithm~\ref{alg:formal_portfolio_optimisation_cvar}} for a single run. Within the numerical experiments, the expectation of the weighted returns is estimated either with CMCI or QMCI and  $\cValueAtRisk_{0.95}$ is estimated using CMCI in order to avoid the propagation of systematic errors that arise when calculating  the expected return and the $\cValueAtRisk$ in the objective function of \eqref{eq:mean-cvar-opt} using QMCI, as discussed in Section~\ref{sec:var-cvar-estimations}. We observe that the objective values estimated using either QMCI or CMCI converge to stable values with an increasing the number of iterations and the curve follow a similar trajectory up to an offset which seems constant. This indicates that QMCI can serve as  an effective subroutine for function estimation within classical optimisation algorithms.  

Figure~\ref{fig:sub_fvssamples} shows the value of the objective function  $f(w,X)$ for 100 independent runs of {Algorithm~\ref{alg:formal_portfolio_optimisation_cvar}} over different number of samples for CMCI and QMCI, respectively. We we can see that the average values of the final objective $\bar{f}(w^*,X)$ obtained by Algorithm~\ref{alg:formal_portfolio_optimisation_cvar} using QMCI and  CMCI converge to consistent stable values as the number of samples increases.

We have performed a Mean-$\cValueAtRisk$ optimisation of asset allocation for two independently distributed Gaussian distributions. As the $\cValueAtRisk$ estimation using QMCI is sensitive to the number of samples as well as the error induced when running QMCI, the $\cValueAtRisk$ was estimated using CMCI during the optimisation procedure, while the expected return is estimated using QMCI. For a future work, we plan to investigate the resources required to evaluate the objective function fully quantumly.

To conclude this Section, we point out a notable difference between the classical and the quantum approach. When using classical Monte Carlo methods for Mean-$\cValueAtRisk$ optimisation, the returns of different portfolio assets are sampled from a given distribution. The same sample is then used to evaluate the objective function (both the expected return and the $\cValueAtRisk$) with different weight vectors (i.e., decision variables) in the classical optimisation algorithm at each optimisation iteration. This notably ensures that the variation of the objective function estimated at different iterations solely arises from the updates of the weights. However with QMCI, we need to re-evaluate the expected return and $\cValueAtRisk$ at each iterative step, and thus the variation of the objectives includes the systematic errors that arise from estimating the objective using QMCI, alongside the variation induced by updating the weight vector. This implies that the classical algorithm with QMCI as a subroutine is only practically realisable when the systematic errors on the objective induced by QMCI are negligible compared to the variation on the objective caused by updating the weight vector. Therefore, this requires us to consider quantum circuits with larger numbers of qubits to sufficiently reduce systematic errors. Determining the precise hardware resources required will be the object of future study. \\

\section{Asset allocation under uncertainty and Mean-Variance optimisation}
\label{sec:asset-allocation-uncertainty}

The objective of this Section is to introduce the problem of determining an optimal asset allocation strategy when market conditions are uncertain; whether this is parameter uncertainty or distributional uncertainty. Indeed, uncertainties on parameters and distributions can be modelled as random variables, random processes, or random fields \cite{Zhang2021modern}, which leads to an SP problem as stated in \eqref{eq:sp-model}. Thus,  QMCI can be a subroutine to aid in solving \eqref{eq:sp-model}.  In the context of mathematical finance, Monte Carlo simulations generate multiple scenarios of future market conditions, and the optimisation problem aims to find the allocation strategy that maximises the expected utility under these scenarios.

\subsection{Framework and definitions} 
A popular formulation of investment optimisation is choosing a point on the Capital Allocation Line (See e.g. Ref.~\cite[page 55]{debrouwer2012maslowian}). This problem is equivalent to choosing an allocation to a risk-free asset with positive return and the market risky portfolio. As a particular case, the reader might want to think of a simplified example where there is only one risky asset. Using this formulation, it is possible to optimise the ``terminal utility'', $ U(V_{T+1}) $, a function of the value $V_{T+1}$ of the portfolio at time $T+1$ (without loss of generality), by choosing the proportion $ \omega_{T} $ of the current wealth $ V_{T} $, at time $T$, to invest in a risky asset  and the remaining in a risk-free asset ($\tilde{\omega}_{T} = 1- \omega_{T} $). We will further assume that no leverage is allowed and hence that $\tilde{\omega}_{T} \ge 0$ and $\omega_{T} \ge 0$. The objective function may then be given as
\begin{align}\label{eq:problem}
    \hat{\omega}_T = \argmax_{\omega_T \in [0,1]} \mathbb{E}_{T}\left[U\left(V_{T+1}\right)\right],
\end{align}
where $ \mathbb{E}_T $ denotes the expectation at time $T$, based on all available information up to that time. The terminal utility $ U(V_{T+1}) $ is a function of the chosen allocation $\omega_T $ and the respective returns of the risky and risk-free assets. The wealth equation is
\begin{equation}
\label{eq:wealth-equation}
\begin{cases}
   V_{T+1} = V_T \times \left(\omega_T e^{r_{T+1} + R^{(f)}_{T}} + \tilde{\omega}_T e^{R^{(f)}_{T}}\right),\\
   \tilde{\omega}_{T}+ \omega_{T} = 1, 
   \end{cases}
\end{equation}

where $ r_{T+1} = X_{T+1} - R^{(f)}_{T} $ represents the ``risk premium'', the excess return of the risky asset over the risk-free rate, $ R^{(f)}_{T} $ (which is the deterministic risk-free rate at time $ T $) and $X_{T+1}$ (which is the return of the risky asset  at the time $T+1$). In practice, we usually consider logreturns instead of raw returns, which can be expressed as  $X_{T+1} = \log\left(\frac{V_{T+1}}{V_T}\right)$.

Equation~\eqref{eq:problem} can be expressed as an integral over all possible future returns, weighted by the predictive distribution of these returns
\begin{align}
\label{eq:optimisation-function-usecase1}
    \hat{\omega}_T = \argmax_{\omega_T \in [0,1]} \int_\Omega U\left(V_T \times (\omega_T e^{r_{T+1} + R^{(f)}_{T}} + (1 - \omega_T) e^{R^{(f)}_{T}})\right) p\left(r_{T+1} | y_T, \Theta\right) \, dr_{T+1},
\end{align}
where $ \Theta $ represents the set of parameters that describe the predictive distribution $ p $ of the excess returns, and $ y_T $ encapsulates all past data up to time $ T $. One more detail is that of the excess market return $ r_{t+1} $ which may be modeled as a linear process
\begin{equation}
 r_{t+1} = \mu_{t+1} + e_{t+1}, \quad t = 1, \ldots, T,
\end{equation}
where $ \mu_{t+1} $ is the expected return at time $t+1$ -- which may be constant or time-varying based on predictors up to time $ t+1 $ -- and $ e_{t+1} $ is a random error term with mean 0 and variance $ \sigma^2 $. This equation deliberately does not specify a distribution for the error term, allowing for a flexible model that can accommodate different forms of distribution uncertainty.

\subsection{Application to the Mean-Var simulation-based portfolio optimisation problem}\label{subsec:mean-var-portfolio-optimisation}

We now consider a simple quadratic utility function which defines the Mean-Var optimisation problem, of form   
\begin{equation}\label{eq:utility_function_mean_var}
U = \E[V] - \lambda \sigma_{V} ^ 2,
\end{equation}
where $\lambda>0$ is a factor that represents the risk aversion (in this simple model it is assumed to be a constant) and $\sigma_{V} =\sqrt{\Var [V]}$ i.e., the volatility of the wealth $V$ of a portfolio of interest. This utility function has a certain theoretical appeal, though we acknowledge that the formulation of the investment problem is rarely setup like this in practice. Firstly, (as argued in \cite{debrouwer2009maslowian} and \cite{debrouwer2012maslowian}) this would limit the  optimisation problem to the time horizon corresponding to the risk-free asset that is under consideration. Secondly, the optimal allocation corresponds to a hypothetical utility function that is not realistic \cite{debrouwer2001fallacy}, and hence impossible to estimate. Thirdly, the assumption that variations (positive and negative) both contribute to risk is not realistic (typically people desire positive returns and aim to avoid negative returns (losses); variance penalises both). However, if an investor were to be rational and uniformly risk averse, this formulation is a reasonable description of reality. From \eqref{eq:utility_function_mean_var} the utility to maximise is thus given by \begin{align}\label{eq:utility_mean_minus_var}
    U(V_{T+1}) &= \E[V_{T+1}] - \lambda \sigma_{V_{T+1}} ^ 2 ,\nonumber \\
       &= \E[V_{T+1}] - \lambda \left( \E[V_{T+1}^2] - \E[V_{T+1}]^2 \right).
\end{align}

\subsubsection{First numerical example: Mean-Var portfolio optimisation with a Gaussian asset}
\label{subsubsec:1st-numerical-exp}

To simulate a realistic example for running numerical experiments, we consider the simplified setting presented in Table~\ref{tab:1st-numerical-exp-params}, where the risk return $r_{T+1}$ is modelled by a Gaussian and we consider a situation with a moderate level of risk aversion. We can expand the objective function given in \eqref{eq:utility_mean_minus_var} by inserting the wealth equation given in \eqref{eq:wealth-equation}, so that the objective function then has an explicit dependence on $\{r_{T+1}, \omega_T, R^{(f)}_{T},\lambda \}$, and if we re-parameterise the equation based on the fact that the exponential of the risk return will be log-normally distributed i.e.,
\begin{equation}
    e^{r_{T+1}} \equiv \rho_{r_{T+1}} \sim  \mathcal{LN}(\mu,\sigma^2),
\label{eq:exp_random_error} 
\end{equation}
we find that the utility to optimise can be expressed as 
\begin{equation}
    \begin{split}
 &U(\rho_{T+1}, \omega_T, R^{(f)}_{T},\lambda) = V_T \left(\omega_Te^{R^{(f)}_{T}}\int_\mathbb{R}  \rho_{t+1} p(\rho_{t+1}) \, d\rho_{t+1} + \tilde{\omega}_T e^{R^{(f)}_{T}} \right) \\
&-\lambda  V_T^{2} \omega_T^{2} e^{2R^{(f)}_{T}} \left(\int_\mathbb{R} \rho_{t+1}^{2}p(\rho_{t+1})\, d\rho_{t+1} -  \left(\int_\mathbb{R} \rho_{t+1} p(\rho_{t+1})\, d\rho_{t+1}\right)^{2}\right).
\end{split}
\end{equation}
The integrals in the expression are equivalent to the expectation value and variance of $\rho_{T+1}$, and these can be calculated using QMCI. The full derivation of this expression is detailed in Appendix~\ref{app:mean-var-calc}. We summarise in Table~\ref{tab:1st-numerical-exp-params} the values chosen to obtain the subsequent numerical results.

\begin{table}[t!]
    \centering
    \begin{tabular}{| p{3cm} || p{3cm} | }
    \hline   
        \textbf{parameter} & \textbf{value / choice}  \\ \hline \hline 
        unit of $T$ & 1 year \\ \hline
        $V_T$  & $100\$ $  \\\hline
        $R^{(f)}_{T}$  & $6\% $  \\\hline
        $\lambda$  & $0.5 $  \\\hline
        $r_{T+1}$  & $\mathcal{N}(\mu,\sigma^2) $  \\ \hline
        $\sigma$  & $0.2 $  \\ \hline
        $\mu$  & $12\%-R^{(f)}_{T} $  \\  \hline
    \end{tabular}
    \caption{Choice of parameters for the numerical example of Mean-Var SBO for portfolio optimisation of Section~\ref{subsec:mean-var-portfolio-optimisation}. The values used and the general approach is similar to \cite{debrouwer2012maslowian}. }
    \label{tab:1st-numerical-exp-params}
\end{table}

\paragraph*{Closed-form solution}
The closed-form solution for the optimisation problem is derived in Appendix~\ref{app:mean-var-calc}. From 
\eqref{eq:minimum_discrete}, the optimal value for the $5$-bit discretised version of the problem with the parameters defined in Table~\ref{tab:1st-numerical-exp-params} is $\hat{\omega}^{\text{disc}}_{T} = 0.0164$. 

\paragraph*{Numerical Experiments}
We now discuss the numerical experiments that were carried out using QMCI as a subroutine to estimate the utility when performing the SBO described in this section. In particular, we explore how the effects of noise impacts the optimisation when performed on the QMCI engine.\footnote{The use case described in this section is the most promising of the two use cases discussed in this paper for a study involving noise, as no enhancement operations are necessary in the circuit, which means the circuits are relatively low depth and do not contain a large number of qubits.} Note that because the purpose of this numerical experiment is to study the performance of QMCI optimisation when performed in both noiseless and noisy regimes, then in contrast to the previous sections we do not focus on comparisons of the performance of QMCI with CMCI. We therefore do not carry out a detailed analysis of errors as was done in previous sections.

The application of the technique of NA-QAE for error mitigation discussed in Section~\ref{subsec:naqae} for mitigating the noise when running QMCI is investigated, under the assumption that the Gaussian noise parameter $k_\mu$ is negligible -- equivalently assuming a depolarising noise model. The numerical experiments were carried out using both Qulac's noiseless state-vector simulator \cite{Suzuki_2021} and a noisy simulator that emulates the result of computations performed on Quantinuum's trapped-ion quantum computer, H1.\footnote{It is known from previous calibration runs on the noisy H1 simulator that $k_\mu$ is negligible for the case of shallow circuits with a small number of qubits, and thus NA-QAE methods should be applicable based on our assumptions.} In general, this study represents the first time that the error-mitigation protocol of NA-QAE has been directly applied to a problem of real-world interest that makes use of QMCI as a subroutine. The notation for this section follows that of Section~\ref{subsec:qae}.

When considering the presence of noise, based on H1 device characteristics as specified in July 2024 \cite{h1csv}, the maximum gate count of circuits (in terms of two-qubit gates) that can be reliably run can be estimated. H1 has a two-qubit gate average infidelity of $1 - \varphi_{2} = 8.830 \times 10^{-4}$. Assuming a depolarising noise model, this corresponds to a depolarising error probability of $p_{\text{dep}} = 1-\frac{4}{3}(8.830 \times 10^{-4}) = 0.999$; ignoring all other sub-dominant sources of noise compared to the two-qubit gate error this means that approximately $850$ two-qubit gate operations can be performed, on average, before there is an error that causes complete decoherence. However, in a future scenario where the infidelity is decreased by approximately 1.7 orders of magnitude (i.e. the infidelity is instead taken as being $1 - \varphi_{2} \sim 1.766 \times 10^{-5}$), then $\sim 42,000$ two-qubit gate operations, on average, can be performed before there is such an error. A 5-qubit circuit which prepares a log-normal distribution for $\rho_{T+1}$ with the parameters given in Table~\ref{tab:1st-numerical-exp-params} contains $\sim 66$ two-qubit gates when compiled to run on H1. Each Grover iteration then adds an additional $\sim 214$ gates. Thus, realistically for the current infidelity, only a few Grover iterates can be performed before errors start to get out of hand.

Information regarding the largest circuits that would be run for given numbers of samples in the noiseless case (assuming the optimal allocation value for $\omega_{T}$) is given in Table~\ref{tab:circuit_m_vals} for both the current infidelity and the future scenario.
\begin{table}[t!]
    \centering
    \resizebox{\textwidth}{!}{
    \begin{tabular}{ |c||c|c|c|c|c|c } 
    \hline 
         \textbf{Samples} & $2000$ & $3000$ & $4000$ & $5000$ & $6000$ \\ \hline \hline
         $m_k$ value (largest circuit) & $16$& $32$ & $32$ & $32$ & $64$ \\ \hline
         two-qubit gates (largest circuit) & $3490$ & $6914$ & $6914$ & $6914$ & $13762$ \\ \hline
         Expected number of depolarising errors (largest circuit, $1 - \varphi_{2} = 8.830 \times 10^{-4}$) & $4$ & $8$ & $8$ & $8$ & $16$ \\ \hline
         Expected number of depolarising errors (largest circuit, $1 - \varphi_{2} = 1.766 \times 10^{-5}$) & $0$ & $0$ & $0$ & $0$ & $0$ \\
     \hline
    \end{tabular}
    }
    \caption{Information about the largest circuits that would be run for given number of samples assuming the optimal allocation. All relevant parameters are defined in Section~\ref{subsec:naqae}.} \label{tab:circuit_m_vals}
\end{table} We can see that by $3000$ and $6000$ samples, i.e. where the largest circuits to run are $m_k=32$ and $m_k=64$, respectively, noise effects would be expected to be significant at the current infidelity. However, in the future scenario, all required computations should be able to be performed coherently.\footnote{Especially considering that for NA-QAE, fewer deep circuits are expected to be run than in the noiseless case i.e., more shots of lower depth circuits will be run.} Thus, in order to be able to meaningfully run NA-QAE (i.e. without noise becoming overwhelming), the level of noise in the simulation was artificially reduced in order to correspond to the future scenario. The number of QMCI samples that were probed are then $\mathcal{S} \in \{ 2000, 3000, 4000, 5000, 6000 \}$.

In order to model the exponential of the risk-return $\rho_{r_{T+1}}$ given in \eqref{eq:exp_random_error}, a 5-qubit log-normal distribution was prepared that was loaded directly in the QMCI engine using the in-built state-preparation library. The QMCI engine was then used to build efficient, low depth circuits that estimate both the expectation value and variance of the random variable $\rho_{r_{T+1}}$ using QMCI.

Within the QMCI engine, an implementation of the MLE-QAE algorithm, optimised based on empirical studies carried out previously to determine the optimal number of shots to run for all $m_k$ values in the EIS, alongside efficient use of all provided samples, was used for the studies. An uniform prior was placed on both $\theta_a$ and $k_\sigma$, where in the latter case the range was from $k_\sigma$ to $2k_\sigma$, based on the estimate $k_{\sigma} = 0.004$ corresponding to $\tilde{p}_{\text{coh}} = 0.992$ derived from \eqref{eq:k_sigma_dep}.

For each total number of samples probed, 100 runs of SBO in the noiseless regime and two runs on the noisy simulator (where one run used NA-QAE to perform error mitigation and the other did not implement any error-mitigation methods), were performed.\footnote{We were only able to perform single runs for each number of samples on the noisy simulator due to the very large computational time required to run the calculations (as it was not possible to parallelise computations when running on the simulator).}

For each run of SBO, the local minimum was determined using the bounded version of Brent's algorithm, which scans the range $\hat{\omega}_T \in [0,1]$ and at each scan point performs QMCI in order to calculate the value of the objective function given in \eqref{eq:utility_mean_minus_var} (by running the expectation value and variance QMCI circuits described previously). We note that Brent's algorithm is not particularly well-suited to optimisation problems in the presence of noise, however given that the aim of this study is to directly compare the performance of the optimisation in both the noiseless and noisy regimes, then we choose the same optimiser in both cases -- with Brent's algorithm chosen for its relative simplicity.

For a given run of QMCI, when calculating the objective function, one consideration that is important is -- for a fixed number of samples -- how best to divide these samples between the expectation value and variance calculations (and thus also to the individual second moment and expectation value calculations that make up the variance calculation), in order to minimise the total error on the final estimate. In order to determine the optimal division of samples, an additional classical minimisation at each scan point, based on the SLSQP algorithm, was performed in order to minimise the theoretical expected asymptotic upper bound on the RMSE of the estimate of the utility for the given number of samples.\footnote{This calculation assumes unbiased estimators -- clearly not the case here for a finite number of samples, but correct in the asymptotic case.} 

\begin{figure}
    \centering
\includegraphics{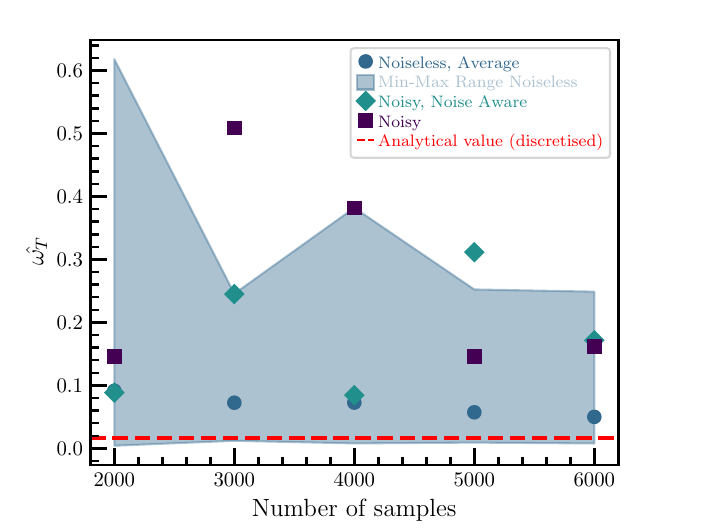}
    \caption{Results of the full SBO for determining the optimal allocation as a function of the number of samples, where the utility function is evaluated at each scan point using the QMCI engine and the respective number of samples. (\textbf{blue}) gives the results using a noiseless simulator, which are averaged across 100 independent runs, and the shaded region represents the range of values observed from those 100 runs. (\textbf{green}) and (\textbf{purple}) give the results using a noisy simulator of Quantinuum's H1 device with a single run in each case, the former with NA-QAE methods applied for error mitigation, and the latter without any error mitigation applied. (\textbf{red}) gives the analytical value for the minimum.}
    \label{fig:average_allocation_noiseless_vs_noisy}
\end{figure}

The results are given in Fig.~\ref{fig:average_allocation_noiseless_vs_noisy}, which demonstrates the optimal allocation as a function of the number of samples for both runs on the noisy emulator (with and without NA-QAE) alongside the average optimal allocation, and the maximum and minimum optimal allocations, observed in the noiseless case. The analytical value for the minimum is also shown. 

Firstly, we note that, unfortunately, due to the fact that only single runs on the noisy emulator were able to be performed, it is difficult to say anything conclusive with regards to the performance of the noisy runs, particularly given the high variance exhibited in the noiseless case. In regards to the comparison of running on the noisy simulator with and without NA-QAE, it may be argued that, on average, the runs performed without applying NA-QAE are slightly further from the true minimum, but this is not always the case (for example the values for $5000$ and $6000$ samples are closer); indeed this can easily be explained by low statistics (a sample size of one). As such, it is difficult to tell if the error mitigation has improved the results. For example, we can see that in both cases the majority of values (four out of five in both cases) are consistent with values that were also observed in the noiseless case.

\begin{figure}
    \centering
\includegraphics{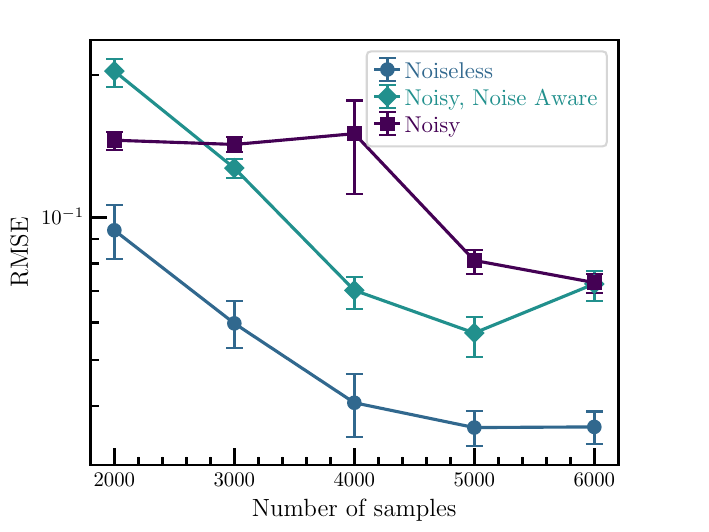}
    \caption{RMSE as a function of the number of samples when the utility function is calculated using the QMCI engine and evaluated at the true analytical minimum $\hat{\omega}^{\text{disc}}_{T}$. (\textbf{blue}) gives the results using a noiseless simulator. (\textbf{green}) and (\textbf{purple}) give the results using a noisy simulator of Quantinuum's H1 device, the former with NA-QAE methods applied for error mitigation, and the latter without any error mitigation applied. All points are averaged across 100 independent runs.}
    \label{fig:rmse_utility_noiseless_vs_noisy}
\end{figure}

In general the issue for this study is that both optimisation and estimation are performed, which makes rigorous comparison difficult. In order to explore the performance in more detail, we now focus on just the estimation part of the routine. We calculated the utility function evaluated at the true optimal $100$ times when run on the noiseless and noisy simulators, in the latter case both with and without using NA-QAE. Figure~\ref{fig:rmse_utility_noiseless_vs_noisy} compares the convergence properties of the RMSE of the estimates as a function of the number of samples. Here we can see that in the noisy case NA-QAE generally improves the convergence, bringing the performance closer to the noiseless case, with four out of five data points having a lower RMSE than when the circuit is run on the noisy emulator without applying error mitigation. This demonstrates the power of the NA-QAE method for this simplified example. As to why there is a discrepancy for 2000 uses, it is likely that the level of noise is so low based on this reduced noise setting (the deepest circuits run have $m_k=16$) that the NA-QAE method does not work effectively -- or this could also be a statistical fluctuation.

Despite being unable to make any conclusive remarks as regards the benefits of applying NA-QAE to the full SBO problem, it is still worth making the point that this is the first time that NA-QAE has been applied to a relevant real-world problem that would make use of QMCI as a subroutine. It is reassuring that the results we obtain are, in general, consistent with results one could also expect to obtain in the noiseless case (even considering the reduced level of noise explored in this study).

Whilst comparisons between the noiseless performance of QMCI and CMCI is not the focus of this section (this is rather the primary focus of the studies in previous sections), if we ignore the noisy runs and just focus on the noiseless case, we can also see that even here the optimisation (on average) is still some way off from converging to the true minimum as would be expected for CMCI, even at $\mathcal{S}=6000$; this is worth discussing briefly and is likely based on the following:

\begin{itemize}
\item The evaluation of the cost function relies on using QMCI to estimate the expectation value and the variance (which is itself composed of separate evaluations of the expectation value and second moment).

\item Because the total number of samples is fixed, the samples specified in the plot are shared between each of these evaluations based on minimising the theoretical total error on the utility as discussed, i.e. each integral does not make use of $\mathcal{S}$ samples but rather some fraction of $\mathcal{S}$.

\item Performing QMCI relies on performing QAE, and previous studies have shown that for smaller numbers of samples, i.e. $\mathcal{S} < 1000$, the convergence properties of the MLE-QAE algorithm differ from the asymptotic behaviour of QAE. The minimisation of the theoretical total error on the utility mentioned above assumes the asymptotic behaviour of the convergence, and this is therefore a source of systematic error. 

\item Each of the integral evaluations in the utility individually contribute their own sources of error. These have different magnitudes depending on the the number of samples assigned to the evaluation (based on running QAE) alongside the specific systematic errors (as described in Section~\ref{sec:systematic-errors-section}) and convergence properties of the particular function being applied to the random variable  (expectation value or second moment in this case). 

\item Because the theoretical error on the second moment calculation, arising from the specific systematic errors and convergence properties of the function applied, is known to be significantly larger than for expectation values \cite{akhalwaya2023modular}, then assuming asymptotic QAE convergence, the majority of samples will be assigned to that calculation. This therefore leads to additional errors on the expectation value calculations that only have a small number of samples assigned, given that the QAE convergences will not necessarily behave as expected.

\item The total error on the utility is thus compounded when these evaluations are combined. As the cost function is then evaluated at each scan point, errors propagate throughout the entire minimisation process, likely resulting in large deviations for the converged optimum.

\end{itemize}

Given these points, it is likely that a very large number of samples (and a larger number of qubits) would be required, such that these sources of error are sufficiently suppressed, to be able to achieve the relevant accuracy to provide quadratic advantage for this particular problem at this scale, even in the noiseless case. We thus conclude that this is likely a problem more suited to running in the future, when the capabilities of quantum computers have been developed further, i.e. when deeper circuits with a larger number of qubits can be run.

\section{Conclusions}
\label{sec:conclusions}

This work can be characterised in the context of hybrid classical-quantum computation where an open question is: how much classical computation should be offloaded to the quantum device? 
For the applications studied, the hybridisation was investigated for the evaluation of the objective function of an optimisation problem, where all or part of the objective function can a priori be estimated quantumly.
Specifically, this paper has thoroughly explored the limitations and bottlenecks associated with using quantum Monte Carlo integration (QMCI) as a subroutine for simulation-based optimisation (SBO) in quantitative finance applications, focusing mainly on the estimation of quantities like Value-at-Risk (VaR) and Conditional-Value-at-Risk (CVaR). 
We identified two key challenges: (i) systematic errors arising from limited resources and (ii) the impact of noise in quantum hardware. 
Through a detailed analyses of these challenges, we demonstrated the critical role that systematic errors play in determining whether QMCI can outperform classical Monte Carlo integration (CMCI), and show that noise-aware quantum amplitude estimation (NA-QAE) can be used to mitigate device noise when estimating the cost function of a SBO.
We also explored the possibility of using QMCI and CMCI together to estimate different terms in a composite cost function.

We identified and provided a detailed description of all the systematic errors.
Then, we analysed the systematic errors in $\ValueAtRisk$ and $\cValueAtRisk$ estimations to show that QMCI not only provides accurate estimations but also offers a quadratic speed-up over classical methods, but only when the systematic errors are sufficiently small. 
We showed that the systematic errors primarily stem from the limited number of qubits used to load probability distributions and the precision of the distribution-loading. 
With these analyses, we then determined an empirical upper bound for the systematic errors to achieve quantum speed-up.
Then, by assuming we can load the distribution with a small state-preparation error and using the empirical upper bound on the error magnitude, we determine the minimum number of qubits required for quantum speed up in CVaR estimations.  This highlights the main bottleneck for the speed-up: the ability to load distributions accurately using a large number of qubits, which remains an open problem in quantum computing.

Additionally, we show that the main limitation for $\ValueAtRisk$ and $\cValueAtRisk$ estimations is the thresholding error, and the size of systematic errors depends on the specific function applied in QMCI.
For instance, the same quantum state used for distribution loading can yield accurate estimations for certain functions, such as mean calculations. 
This insight led us to propose the use of QMCI and CMCI in tandem, where QMCI estimates terms in a composite cost function that have small associated systematic errors, while CMCI handles other terms. 
This hybrid approach demonstrates that quantum and classical methods can be effectively integrated to leverage the strengths of each.

Lastly, we examined the effects of noise in quantum hardware on SBO applications. 
While NA-QAE appears to improve results for individual cost function estimations by bringing them closer to the noiseless case, the overall impact on the full SBO problem remains inconclusive. 
This is primarily due to the complexity of the optimisation process, which requires more extensive testing on quantum hardware. 
Nevertheless, for estimation tasks, NA-QAE shows promise as an effective error mitigation technique.

In summary, while QMCI has the potential to offer significant computational advantages over classical methods for financial applications, there remain several open challenges that must be addressed for these advantages to be fully realised. 
Chief among these challenges is the need for efficient and accurate methods to load probability distributions using a large number of qubits, as well as managing the impact of noise in current quantum hardware. 
Despite these challenges, the hybrid QMCI-CMCI approach and NA-QAE represent promising strategies for overcoming some of these limitations, laying the groundwork for further exploration in the intersection of quantum computing and quantitative finance.

It is known that underlying assets being modelled by Gaussian and log-normal distributions is unrealistic, as discussed in Appendix~\ref{app:stable-distrib-levy}. Thus a possible and important extension to this work is analyse more realistic distributions, such as heavy-tailed distributions -- see Appendix~\ref{app:levy_distribution_section} for our initial analyses of the systematic errors associated with loading  L\'evy distributions on quantum computers.
Finally, as another promising extension of our work, the framework discussed in this work can be further generalised and we leave this as an open problem for the community. For instance, one might consider a more complex problem such as ``trade-off optimisation'' with different risk measures -- see Appendix~\ref{app:cdar_optimisaton}.

\section*{Acknowledgements}

We thank M.~Spranger for his work that helped form the background to this paper, and for all of the fruitful discussions
we had along the way. GK would like to acknowledge that this work has been partially supported by project MIS 5154714 of the National Recovery and Resilience Plan Greece 2.0 funded by the European Union under the NextGenerationEU Program. GK further acknowledges support of the Czech Science Foundation (2307947S).

\subsection*{Disclaimer}
This paper was prepared for information purposes
and is not a product of HSBC Bank Plc. or its affiliates.
Neither HSBC Bank Plc. nor any of its affiliates make
any explicit or implied representation or warranty and
none of them accept any liability in connection with
this paper, including, but not limited to, the completeness,
accuracy, reliability of information contained herein and
the potential legal, compliance, tax or accounting effects
thereof. Copyright HSBC Group 2024.

\normalem

\bibliography{bibliography}
\bibliographystyle{IEEEtran}

\appendix

\newpage

\section{Analytical derivation of the utility function and minimum for the Mean-Var portfolio problem}
\label{app:mean-var-calc}
We derive in this Section the analytical expression of the utility function to be minimised and its analytical minimum for the Mean-Var portfolio problem discussed in Section~\ref{subsec:mean-var-portfolio-optimisation}. 
\subsection{Simplifying the utility function}

Starting from \eqref{eq:utility_mean_minus_var} and substituting \eqref{eq:wealth-equation}, we have 
\begin{equation}
    \begin{split}
       U(r_{T+1}, \omega_T, R^{(f)}_{T},\lambda) &= \E\left[V_T\times \left(\omega_T e^{r_{T+1} + R^{(f)}_{T}} + \tilde{\omega}_T e^{R^{(f)}_{T}}\right)\right] - \lambda \left( \E\left[V_T^2\times \left(\omega_T e^{r_{T+1} + R^{(f)}_{T}} + \tilde{\omega}_T e^{R^{(f)}_{T}}\right)^2\right] \right.  \\
       &\left. - \E\left[V_T \times \left(\omega_T e^{r_{T+1} + R^{(f)}_{T}} + \tilde{\omega}_T e^{R^{(f)}_{T}}\right)\right]^2 \right). 
    \end{split}
\end{equation}

Given that we can take $R^{(f)}_{T}$ as a constant for a single time period, this can be written as the following integral to be optimised

\begin{equation}
    \begin{split}\label{eq:utility}
&-U(r_{T+1}, \omega_T, R^{(f)}_{T},\lambda) = - \left[ V_T \omega_T e^{R^{(f)}_{T}} \int_\R  e^{r_{T+1} } p(r_{T+1}) \, dr_{T+1} + V_T  \tilde{\omega}_T e^{R^{(f)}_{T}} \right] \\
&+ \lambda  V_T^2 \omega_T^2  e^{2R^{(f)}_{T}} \left(  \int_\R e^{2r_{T+1}} p(r_{T+1})\, dr_{T+1}-\left(\int_\R e^{r_{T+1}} p(r_{T+1}) \, dr_{T+1}\right)^2 \right).
\end{split}
\end{equation}

To evaluate this utility function we thus just need to estimate the following integrals
\begin{equation}
  \mathcal{I}_1=  \int_\R e^{r_{T+1}} p(r_{T+1}) \, dr_{T+1}, \qquad 
    \mathcal{I}_2=\int_\R e^{2r_{T+1}} p(r_{T+1}) dr_{T+1},
\end{equation}
using MCI. Alternatively, if we take the variable of integration (i.e., the random variable) as $\rho_{t+1} = e^{r_{T+1}}$ instead, then we can write the utility to calculate using MCI as 

\begin{equation}
    \begin{split}
&-U(\rho_{T+1}, \omega_T, R^{(f)}_{T},\lambda) = - V_T\omega_Te^{R^{(f)}_{T}}\int_\mathbb{R}  \rho_{t+1} p(\rho_{t+1}) \, d\rho_{t+1} + V_T\tilde{\omega}_T e^{R^{(f)}_{T}}  \\
&+  \lambda V_T^{2} \omega_T^{2} e^{2R^{(f)}_{T}}   \left(\int_\mathbb{R} (\rho_{t+1})^{2}p(\rho_{t+1})\, d\rho_{t+1} \right.   -  \left. \left(\int_\mathbb{R} \rho_{t+1} p(\rho_{t+1})\, d\rho_{t+1}\right)^{2}\right),
\end{split}
\end{equation}
which amounts to calculating the mean and variance of $\rho_{t+1}$ using MCI. To evaluate this utility function we thus just need to estimate the following integrals
\begin{equation}
  \mathcal{I}_3=  \int_\R \rho_{t+1} p(\rho_{t+1}) \, d\rho_{t+1}, \qquad 
    \mathcal{I}_4=\int_\R \rho_{t+1}^{2} p(\rho_{t+1}) d\rho_{t+1},
\end{equation}
using MCI. Due to the optimisation being found to be more stable based on the approach of modelling the random variable as $\rho_{t+1}$,\footnote{It is more stable because the pre-factors in front of the various integrals to compute are generally smaller, so that fluctuations in the output of the estimation of the integrals are ``amplified'' less in the final calculation. In practice this amounts to a `smoother' optimisation trajectory curve.} this is what is thus adopted.

\subsection{Optimising the utility function for a Gaussian random variable}

To find the analytical value for the optimum we start from the expression obtained in \eqref{eq:utility}. If we take returns as being Gaussian distributed $r_{T+1}\sim \mathcal{N}(\mu, \sigma^2)$ then this expression simplifies to

\begin{equation}
    \begin{split}
&-U(r_{T+1}, \omega_T, R^{(f)}_{T},\lambda) = - \left[ V_T \omega_T e^{R^{(f)}_{T}}e^{\mu+\frac{\sigma^2}{2}} + V_T  \tilde{\omega}_T e^{R^{(f)}_{T}} \right]  \\
&+ \lambda \left[V_T^2 \omega_T^2  e^{2R^{(f)}_{T}}e^{2(\mu+\sigma^2)} + 2V_T^2\omega_T\tilde{\omega}_T e^{2R^{(f)}_{T}}e^{\mu+\frac{\sigma^2}{2}}   + V_T^{2} \tilde{\omega}_T^2 e^{2R^{(f)}_{T}} - \left( V_T \omega_T e^{R^{(f)}_{T}}e^{\mu+\frac{\sigma^2}{2}} + V_T  \tilde{\omega}_T e^{R^{(f)}_{T}}\right)^2 \right],
\end{split}
\end{equation}

Then differentiating we have

\begin{align}
&\frac{\partial (-U)}{\partial\omega_{T}} =  -V_T e^{R^{(f)}_{T}}e^{\mu+\frac{\sigma^2}{2}} + V_T e^{R^{(f)}_{T}} \nonumber \\
& +\lambda \left[2   e^{2R^{(f)}_{T}}e^{2(\mu + \sigma^{2})} V_T^2\omega_T + 2 e^{2R^{(f)}_{T}}e^{\mu+\frac{\sigma^2}{2}} V_T^2(1-\omega_T) \, \right. \nonumber \\
& \left. - 2 e^{2R^{(f)}_{T}} V_T^{2}(1-\omega_{T}) - 2e^{R^{(f)}_{T}}e^{\mu+\frac{\sigma^2}{2}} V_T^2\omega_{T} \, \right. \nonumber \\
& \left. -2\left(-e^{R^{(f)}_{T}}V_T+e^{R^{(f)}_{T}}e^{\mu+\frac{\sigma^2}{2}}V_T\right)\left(e^{R^{(f)}_{T}}V_T(1-\omega_{T}) + e^{R^{(f)}_{T}}e^{\mu+\frac{\sigma^2}{2}} V_T\omega_{T}\right)
\right],
\end{align}
Setting this equal to zero and solving gives the minimum as \begin{equation}\label{eq:minimum}
    \hat{\omega}_T = \frac{e^{-R^{(f)}_{T}-2\mu-\sigma^2}(e^{\mu+\frac{\sigma^2}{2}}-1)}{2V_T\lambda(e^{\sigma^2}-1)}.
\end{equation}

We know however that when we are actually evaluating the minimum using the QMCI engine, as we only have a limited number of qubits available -- which we denote as $n_q$ -- we are not actually calculating the integrals based on a pseudo-continuous distribution (as one could effectively do classically using as many bits as required for precision) but are instead calculating using a discrete approximation to the distribution. Thus if we would like to evaluate the performance of the optimisation using the QMCI engine, we should compare to a discretised version of the problem.

\subsection{Optimising the utility function using a discrete random variable}

If we model this discrete distribution over $N = 2^{n_{q}}$ points as a series of weighted delta functions treated as a continuous PDF such that \begin{equation}
p(r_{T+1}) \approx \sum_{i=0}^{N-1}\omega^{i}(\mu,\sigma^{2})\delta(r_{T+1}- r_{T+1}^{i}),
\end{equation}
then we thus have
\begin{align}
&\mathcal{I}_1^{\text{disc}}=  \int_\R e^{r_{T+1}} \sum_{i=0}^{N-1}\omega^{i}(\mu,\sigma^{2})\delta(r_{T+1}- r_{T+1}^{i}) \, dr_{T+1} = \sum_{i=0}^{N-1}\omega^{i}(\mu,\sigma^{2})e^{r^{i}_{T+1}}, \nonumber \\
&\mathcal{I}_2^{\text{disc}}=\int_\R e^{2r_{T+1}} \sum_{i=0}^{N-1}\omega^{i}(\mu,\sigma^{2})\delta(r_{T+1}- r_{T+1}^{i}) \, dr_{T+1} = \sum_{i=0}^{N-1}\omega^{i}(\mu,\sigma^{2})e^{2r^{i}_{T+1}},
\end{align}
and then following the same methodology as earlier, we get the discrete approximation to the minimum as

\begin{equation}\label{eq:minimum_discrete}
    \hat{\omega}_T^{\text{disc}} = -\frac{e^{-R^{(f)}_{T}}\left[\sum_{i=0}^{N-1}\omega^{i}(\mu,\sigma^{2})e^{r^{i}_{T+1}}-1\right]}{2V_T\lambda\left[\left(\sum_{i=0}^{N-1}\omega^{i}(\mu,\sigma^{2})e^{r^{i}_{T+1}}\right)^2 - \sum_{i=0}^{N-1}\omega^{i}(\mu,\sigma^{2})e^{2r^{i}_{T+1}}\right]}.
\end{equation}

\section{Stable distributions and extended interesting distributions} \label{app:stable-distrib-levy}
A number of works in the quantum finance literature have focused on examples involving Gaussian random variables. As an effort to bring quantum finance closer to more realistic problems, we discuss here stable distributions and other interesting distributions which could be investigated in future studies.

\subsection{Stable distributions} In formulation \eqref{eq:mean-cvar-opt} for the Mean-$\cValueAtRisk$ optimisation problem, the  return of the portfolio $R$  is a random variable. In Markowitz theory, it is traditionally assumed that the returns follow a Gaussian or log-normal distribution. In practice, returns have fat tails and are not distributed as such\cite{Lehoczky2001,fukunaga2018universallevysstablelaw}, hence one needs to investigate more general distributions. An interesting family of distributions are the stable distributions which aim to verify the following definition

\begin{definition}[Stability of a random variable]
    Consider a random variable $X$ and a set $\{ X_1, \ldots, X_n \}$ where each $X_i$ are i.i.d. copies of $X$. Then, $X$ is said to be \emph{stable} iff for each $n>1$, there exists scaling factor $c_n>0$ and a shift factor $d_n$ such that
    \begin{align}
        \sum_{i=1}^n X_i = c_nX + d_n.
    \end{align}
\end{definition}
Returns of financial assets inherently have the property of stability since the return of asset $i$ at time $[t_j, t_{j+T}]$, is given as 
$$
R_i(t_j) = \frac{V(t_{j+T}) - V(t_j) }{V(t_j)} \approx \log V(t_{j+T}) - \log V(t_j),
$$
that is continuous compounded return. Therefore 
$$
R_{i}([t_1\to t_k]):=\sum_{\ell=1}^k R_i(t_\ell) \approx \log V(t_{k+1}) - \log V(t_1)  ,
$$
where $R_{i}([t_1\to t_k])$ represents the total return from $t_1$ to $t_{k+1}$, and is equivalent to the sum of the individual returns over each time period from $1$ to $k$.

Among the stable distributions, $\alpha$-stable distributions offer the most realistic models \cite{Krezolek2015} in mathematical finance.  An $\alpha$-stable distribution is a family of distributions depending on four parameters, specified by stability index $\alpha \in (0,2]$, skewness $\beta$, scale $\gamma$, and location $\delta$. Gaussian distributions form a special case where $\alpha=2, \beta=0$ \cite{bouchaud2000theory}. Interestingly, empirical evidence suggests that $\alpha$-stable distributions are comparatively more risk adverse. The difficulty with $\alpha$-stable distributions lie in the fact that direct sampling from the characteristic function is not possible. Fourier inversion formula and L\'evy’s characterisation theorem are required in order to invert these distributions and obtain the density or distribution function. There exist classical methods to try to overcome this problem such as Devoye's method \cite{Devroye1986-ro} or Br\"uck's generative neural network \cite{bruck2024generative}. It remains an open problem to extend these methods to the quantum setting.

\subsection{Other interesting distributions}
We now list some other distributions which have not to our knowledge been studied in the quantum literature and would be of interest to use for quantum algorithms that apply to more realistic use cases.

\begin{itemize}
    \item \textbf{Shifted log-normal distribution}.
The shifted log-normal distribution extends the
Gaussian distribution via an additional parameter controlling skewness \cite{johnson1995continuous}. The PDF is given by
\begin{equation}
f(x)=\frac{1}{(x-\theta) \sigma \sqrt{2 \pi}} \exp \left(-\frac{(\ln (x-\theta)-\mu)^2}{2 \sigma^2}\right),
\end{equation}
for $x>\theta$. 
\item \textbf{Generalised Gaussian distribution}.
The generalised Gaussian distribution extends the Gaussian distribution via an additional parameter
controlling kurtosis \cite{Nadarajah2005}.\\
\end{itemize}

The shifted log-normal and generalised Gaussian distributions are of particular interest in financial modeling due to their ability to model and represent non-trivial, non-Gaussian characteristics that are frequently observed in financial data. The shifted log-normal distribution skewness parameter $\theta$ allows for the modeling of asymmetric distributions with a lower bound. This is particularly relevant in financial modelling applications where certain variables, such as asset prices or returns, are constrained by a minimum value (such as a fundamental price or a regulatory requirement). By incorporating this shift, the distribution can more accurately reflect the empirical properties of the underlying data, leading to enhanced model realism and improved risk assessment. 

On the other hand, the generalised Gaussian distribution extends the traditional Gaussian framework by including a parameter to control the kurtosis of the distribution. This additional flexibility enables the distribution to adapt to the leptokurtic nature of financial returns (characterised by heavy tails and sharp peaks). This is crucial since the ability to capture and model excess kurtosis is crucial for accurate risk estimation, particularly in the context of extreme events and tail-risk management. By employing the generalised Gaussian distribution, one may obtain a more precise quantification of the probabilities associated with rare, high-impact events, facilitating robust stress testing and capital allocation decisions.

\section{Error analysis for the L\'evy distribution}\label{app:levy_distribution_section}

It is an established fact that, at least over small time-periods, the distributions of returns of financial assets are heavy-tailed compared to the Gaussian distribution \cite{bouchaud2000theory}. This implies that the Gaussian distribution significantly under-estimates extreme returns. Since financial returns are, in essence, convolutions of returns on smaller time scales, it is sensible to consider L\'evy distributions \cite{Nolan2020,Elmerraji2021,Gong2021}\footnote{and more generally stable distributions, see Appendix~\ref{app:stable-distrib-levy}} over smaller time-periods as they are stable under addition. They are able to embody characteristics of market data such as heavy tails and asymmetry which are crucial for modeling financial risks and returns over different time scales. The PDF of the Lévy distribution, denoted by $\mathcal{L}(\mu,c)$, is defined for $x \geq \mu$ where $\mu$ is the location parameter and $c$ is the scale parameter. Its PDF is
$$
\mathcal{L}(x,\mu,c)=\sqrt{\frac{c}{2 \pi}} \frac{e^{-\frac{c}{2(x-\mu)}}}{(x-\mu)^{3 / 2}},
$$
highlighting its heavy-tailed nature which results in infinite variance and, under certain conditions, infinite expectation. 

\begin{figure}[!htb]
    \centering
    \includegraphics{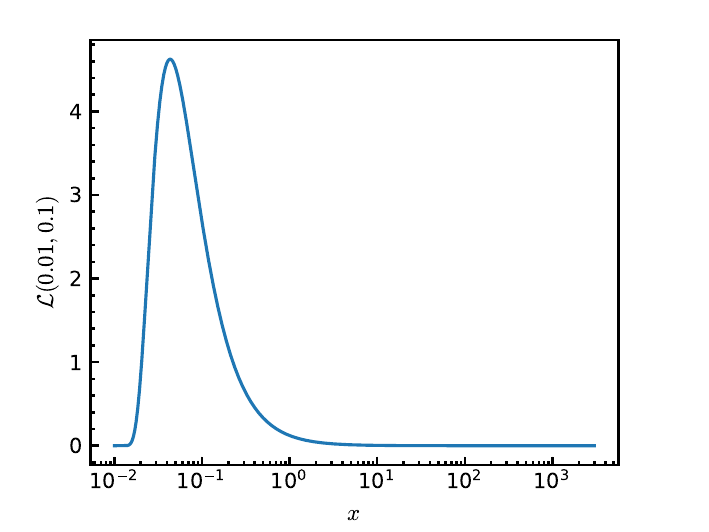}
    \caption{Trained L\'evy distribution with $\mu=0.01$ and $c=0.1$ in the range $\left[0.01, 3000\right]$.}
    \label{fig:levy_example}
\end{figure}

As $x$ approaches positive or negative infinity, the functional form of the Lévy distribution approximates $\frac{\mu A_\pm^\mu}{|x|^{\mu+1}}$, where $0 < \mu < 2$, with $\mu$ being the stability index or characteristic exponent, and $A_{\pm}$ represents the tail amplitudes which determine the decay scale of the distribution's tails. Given the inherent asymmetry, quantified by the asymmetry parameter $\beta = \frac{A_+^\mu - A_-^\mu}{A_+^\mu + A_-^\mu}$, Lévy distributions are particularly suited for modeling financial instruments or portfolios where risk assessment is critical \cite{papapantoleon2008introduction}. For instance, in risk management, the heavy-tailed feature of the Lévy distribution implies that risk measures like $\ValueAtRisk$ and $\cValueAtRisk$ can reach particularly high values, reflecting significant potential losses in adverse market conditions. Due to these properties, Lévy distributions are often truncated in practical applications to manage the implications of their infinite support, focusing instead on a finite range of practical interest, which simplifies both the theoretical analysis and numerical computations involved in financial risk assessments. 

\begin{figure}[!t]
    \centering
    \includegraphics{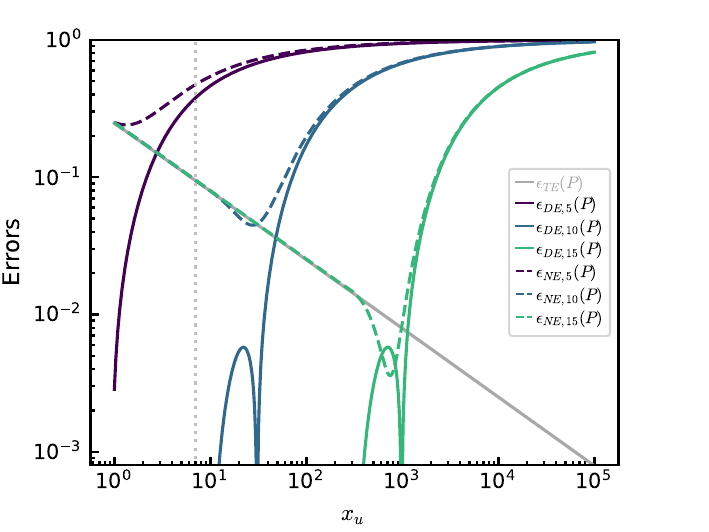}
    \caption{Truncation, discretisation, and normalisation errors for the L\'evy distribution $\mathcal{L}(0.01, 0.1)$ as a function of $x_{u}$ for 5-, 10-, and 15-qubit circuits.
    The truncation error (as expected) gets smaller with increasing $x_{u}$.
    The discretisation and normalisation errors get smaller as the number of qubits is increased, but increase with $x_{u}$ for a fixed number of qubits.
    The vertical dashed silver line represents the $x_{u}$ choice used for the trained circuit represented in Fig.~\ref{fig:trained_circuit}.
    }
    \label{fig:discretization_error_2}
\end{figure}

We subsequently study a L\'evy distribution with a truncated support range $\left[\mu, x_{u}\right)$ and as in Section~\ref{subsec:param-quantum-circuit}, the circuit representing this distribution is trained using a hardware-efficient ansatz. The trained distribution is given in Fig.~\ref{fig:trained_circuit}. The aim of this Section is to investigate the impact in terms of quantum resources of implementing a distribution that is more suitable for real-life applications.

\subsection{How to choose the upper bound $x_{u}$}
For practical reasons (the limited number of qubits available for quantum hardware and the lack of efficient state-preparation techniques for a large number of qubits), we are restricted to using circuits with a small number of qubits. This poses an extra challenge for the L\'evy distribution due to its heavy tail. To understand this, we carry out an analysis of the systematic errors, and thus we start by defining these errors.

\subsubsection{Error definitions}
In this Section, we provide expressions for the systematic errors when the function applied is the constant 1, which is equivalent to integrating the PDF (without a function applied).
For this, we first write the truncated cumulative probability:
\begin{equation}
    P_{T} = \int_{x_{\ell}-\Delta/2}^{x_{u}+\Delta/2}\mathcal{L}(\mu,c)dx,
\end{equation}
and the truncation error:
\begin{equation}
   \epsilon_{TE}(P) = 1 - P_{T}.
\end{equation}
The discretised probability is given by
\begin{equation}
    P_{D, n} = \sum_{x=x_{\ell}}^{x_{u}} \mathcal{L}(x,\mu,c)\Delta,
\end{equation}
where $2^{n}$ is the number of terms in the summation and $n$ is the number of qubits used to load the distribution.
Finally, we write the discretisation error as
\begin{equation}
    \epsilon_{DE, n}(P) = \abs{P_{T} - P_{D}},
\end{equation} 
and the normalisation error as
\begin{equation}
    \epsilon_{NE, n}(P) = 1 - P_{D, n}.
\end{equation}

\begin{figure}[!t]
    \centering
    \includegraphics{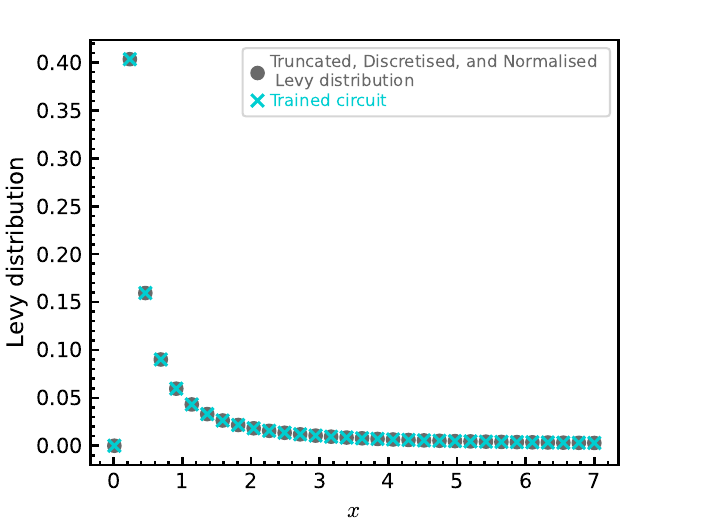}
    \caption{$\mathcal{L}(0.01, 0.1)$ distribution prepared by the trained circuit.
    The distribution obtained from trained circuit has a small state-preparation error and is qualitatively a good approximation ($L_{\infty} \sim 10^{-7}$) of the training target i.e. the truncated, discretised, and normalised L\'evy distribution.}
    \label{fig:trained_circuit}
\end{figure}

\subsubsection{The trade-off between truncation and discretisation errors}
In Fig.~\ref{fig:discretization_error_2}, we present the truncation, discretisation, and normalisation errors for the L\'evy distribution $\mathcal{L}(0.01, 0.1)$ as a function of $x_{u}$ for 5-, 10-, and 15-qubit circuits. Figure~\ref{fig:discretization_error_2} shows that we need to choose a large $x_u$ value ($\sim 10^3$) to have $P_{T} \sim 0.99$ which is equal to $1 - \epsilon_{TE}(P)$).
However, as we can also see in Figs.~\ref{fig:levy_example}-\ref{fig:discretization_error_2}, the majority of the probability density is still in the smaller range $[\mu, 1]$.
Therefore, choosing a large $x_u$ value will lead to considerable discretisation error when we have a limited number of support points.
Fig.~\ref{fig:discretization_error_2} also shows the discretisation and normalisation errors when we use $2^5$, $2^{10}$, and $2^{15}$ support points for $\mathcal{L}(0.01, 0.1)$.
We can see that, in general, there is a trade-off between the truncation and discretisation errors.
For the 5-qubit case, choosing $x_{u} = 7$ gives us $\epsilon_{T} \sim 0.1$, which is shown by the vertical dashed line, and this is what we chose to train for Section~\ref{subsec:levy_trained_dist}.

Finally, in Fig.~\ref{fig:discretization_error_2}, we can see that a larger number of support points lead to lower discretisation errors, as expected.
Interestingly, it is not monotonically increasing, and this means one can go up to $P_{T} \sim 0.95$ for a small discretisation error at certain ``sweet spots''.
However, training or loading a distribution with a small state-preparation error for circuits with a large number of qubits is a challenging and interesting question that is left for future study.

\subsection{5-qubit trained circuit for the L\'evy distribution}\label{subsec:levy_trained_dist}
Finally, for demonstration purposes, we trained a 5-qubit circuit to load the truncated, discretised, and normalised $\mathcal{L}(0.01, 0.1)$ distribution.
We use the HWE ansatz. 
Fig.~\ref{fig:trained_circuit} shows the distribution loaded by such a circuit and the corresponding truncated, discretised, and normalised $\mathcal{L}(0.01, 0.1)$ distribution.
Though the trained circuit has a small state-preparation error and is a good approximation of the training target, it is not a good approximation for the continuous distribution shown in Fig.~\ref{fig:levy_example} (see the silver dashed-line in Fig.~\ref{fig:discretization_error_2} for the sizes of the systematic errors).

\section{Portfolio optimisation with Conditional-Drawdown-at-Risk }\label{app:cdar_optimisaton}

The``trade-off optimisation'' problem, framed within a SBO portfolio-optimisation framework, seeks to balance the return expectations against risk considerations, quantified through the Conditional-Drawdown-at-Risk (CDaR) measure \cite{Chekhlov2004}.
The problem can be formulated as

\begin{maxi}
{w \in \mathcal{W}}{  \mathbb{E}[X] - \lambda \operatorname{CDaR}(w).} 
{}{} \label{form:opt2}
\end{maxi}
Specifically, CDaR is defined as
$$
\operatorname{CDaR}({w}, \beta)=\frac{1}{1-\beta} \int_{D(w, r, t) \geq \alpha(w)} D(w, r, t) p(r(t)) \, d r(t),
$$
where $w\in \mathcal{W}\subset \mathbb{R}^n$ is the vector of portfolio weights,  $r$ is the vector of cumulative asset returns (daily), with probability mass function $p(r(t))$. Furthermore, $D(w, r, t)=\max _{r<t}\left(w^T r(\tau)-{w}^T {r}(t)\right)$ is the drawdown of the portfolio, $\alpha$ is the portfolio drawdown (DaR) with confidence $\beta$ and $0\leq \tau \leq t$.

Here, the objective function is crafted to maximise the expected returns $\mathbb{E}[X] $, adjusted by a risk aversion parameter $\lambda$ that scales the impact of the CDaR. The CDaR, is a risk measure that evaluates the worst expected loss under normal market conditions over a specified period within a given confidence interval
and the risk aversion parameter $\lambda$ is a scalar that quantifies the decision-maker's risk tolerance, influencing the trade-off between risk and return \cite{Chekhlov2000}. Other risk measures, for example Entropic-Value-at-Risk can also be considered for our approach. Nevertheless, we leave this for future work.
\end{document}